\documentclass[pra, reprint,floatfix,superscriptaddress]{revtex4-2}
\usepackage{amsmath, amssymb, mathtools, bm, bbold, booktabs, pgf} 
\usepackage[export]{adjustbox}
\usepackage{float}
\usepackage[colorlinks=true,citecolor=blue,linkcolor=blue,urlcolor=blue, breaklinks]{hyperref}
\usepackage[mathscr]{eucal}
\usepackage{enumitem}

\begin{document}

\title{Multiparticle quantum walks for distinguishing hard graphs}

\author{Sachin Kasture, Shaheen Acheche, Loic Henriet, Louis-Paul Henry}
\affiliation{PASQAL SAS, 2 av. Augustin Fresnel, 91120 Palaiseau, France}

\date{\today}

\begin{abstract}
Quantum random walks have been shown to be powerful quantum algorithms for certain tasks on graphs like database searching, quantum simulations etc. In this work we focus on its applications for the graph isomorphism problem. In particular we look at how we can compare multi-particle quantum walks and well known classical WL tests and how quantum walks can be used to distinguish hard graphs like CFI graphs which k-WL tests cannot distinguish. We provide theoretical proofs and empirical results to show that a k-QW with input superposition states distinguishes k-CFI graphs. In addition we also prove that a k-1 QW with localized input states distinguishes k-CFI graphs. We also prove some additional results about strongly regular graphs (SRGs).	
\end{abstract}\maketitle

\maketitle

%======   INTRO =========

\section{Introduction}

Quantum walks (QW) have been studied in great detail for their non-classical dynamics stemming from quantum interference which leads to different propagation characteristics than a classical random walk. Besides being shown to be a model for universal quantum computation ~\cite{childs2009}, algorithms based on QW have shown to have an advantage over classical walks for problems like database search ~\cite{childs2004} and quantum simulations~\cite{childs2012,mohseni2008}. They have been shown to demonstrate exponential advantage over any classical algorithm for hitting times on a welded tree graph~\cite{childs2003}, hypercube graphs~\cite{krovi2006} and for more general hierarchical graphs ~\cite{balasubramanian2023exponential} because of an effective lower dimensional subspace over which they propagate. QW with time-modulated hopping rates~\cite{schulz2024} have been combined with quantum annealing schedules to solve optimization problems.\\
QW have also been shown to be very promising candidates for the graph isomorphism (GI) problem~\cite{gamble2010,Rudinger2013} and have been used to distinguish hard graphs like strongly regular graphs (SRGs). Different distinguishing abilities have been observed on the basis of choice of interactions and particle type (boson/fermion). In particular interaction has been shown to be important for distinguishing SRGs. Given the applications of GI in fields like pattern recognition~\cite{conte2004}, chemical database search~\cite{merkys2023}, electronic circuit design~\cite{ABIAD2020} etc, QRW provide a powerful alternative to classical approaches for these applications. \\
GI problem has been studied using classical graph theory for several decades. The most powerful known algorithm for GI~\cite{babai2016} runs in quasi-polynomial time ($exp((log (n)^{O(1)})$), where $n$ is the number of nodes in the graph. However, on the practical side, Weisfeiler-Lehman (WL) tests~\cite{weisfeiler1968reduction} are popular classical algorithms used to distinguish graphs. These algorithms apply colors to nodes and goes through several iterations till the nodes reach stable coloring. If the histogram of colors for two graphs are different, then they are non-isomorphic. Higher dimensional k-WL tests~\cite{grohe2015pebble,grohe2017descriptive} are more powerful variants which generate effective higher dimensional graphs with nodes labeled with k-tuples followed by node-coloring iterations.\\
Quantum walks have been implemented using various hardware platforms like NMR~\cite{du2003}, neutral atoms~\cite{karski2009}, trapped ions~\cite{schmitz2009} and integrated photonics ~\cite{peruzzo2010,qiang2021}.
Neutral atom platforms are particularly suited to QRWs, as hardcore-boson random walks can be implemented, in the form of a XY model, using dipole-dipole interaction between two Rydberg states~\cite{browaeys2020,struck2013,barredo2015,de_leseleuc2019,dalyac2024graph}. In that case, having more than 1 particle at a location is forbidden, corresponding to an infinite energy cost.\\  
\begin{figure*}[t!]\centering
\includegraphics[scale=0.3]{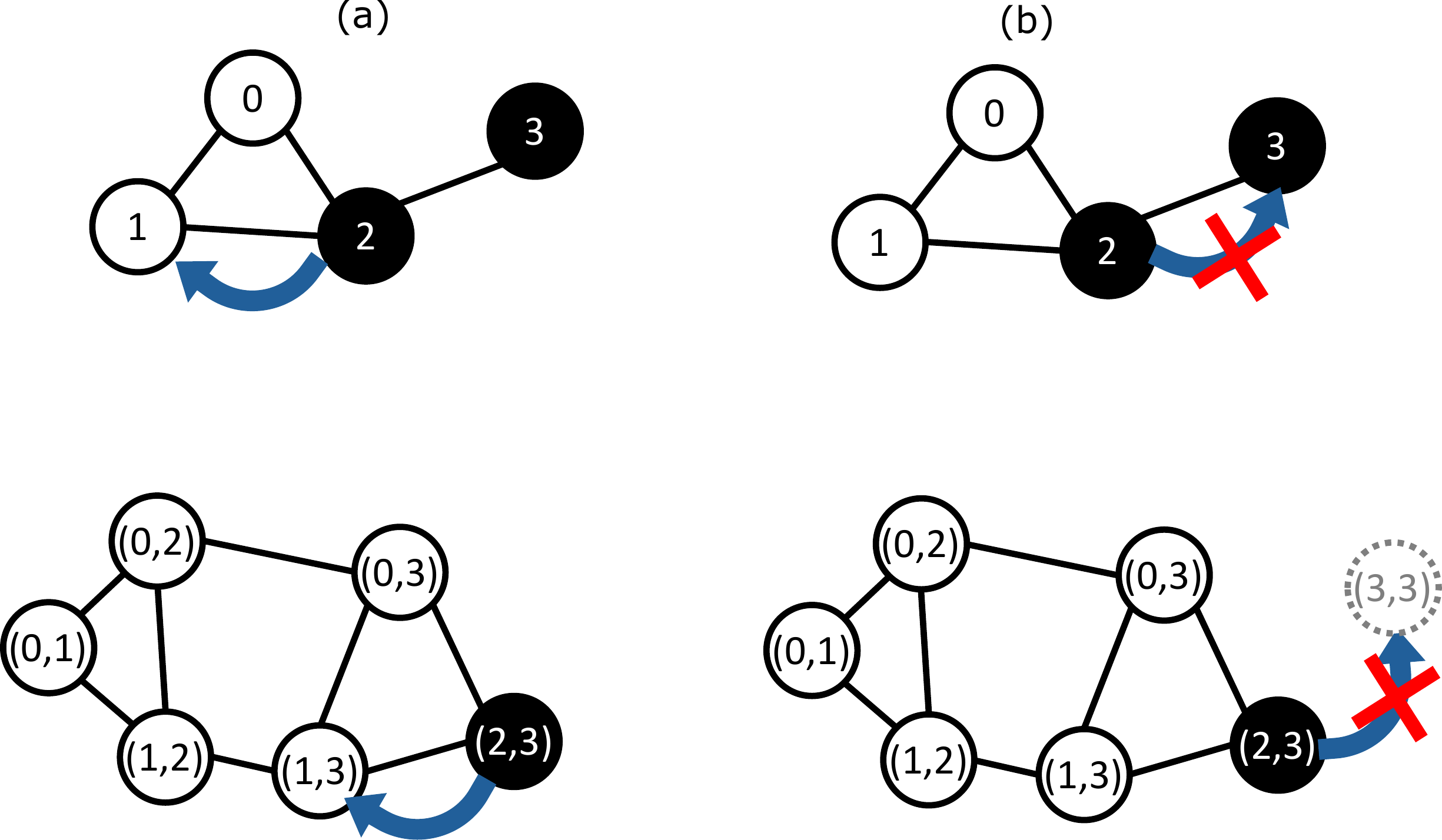}
\caption{Figure illustrates the prohibition of double occupancy in the XY evolution. The up and down panels display the original graph and the 2-particle occupation graphs as discussed in text.(a) shows one of the allowed transitions. The 2 particles begin with node positions 2 and 3 and end up at 1 and 3. In the occupation graph picture, this corresponds to transition from 23 to 13.(b) shows a forbidden transition. Particles beginning with positions 2 and 3 cannot end up at 3 together. In the occupation graph picture, this corresponds to 33 being an isolated node.}
\label{double_occupation}
\end{figure*}
In this work, we study how k-particle QW (k-QW) perform on families of hard to distinguish graphs known as CFI graphs ~\cite{cai1992,morris2020weisfeiler}. These graphs are build so that, for a given integer $k\geq 2$, k-CFI graphs cannot be distinguished by a k-WL test. In order to successfully distinguish them, one needs to use (k+1)-WL test or higher. Here we show both theoretically and empirically that k-QW distinguish k-CFI graphs, when the input state is a uniform superposition. While it is experimentally challenging to create uniform superposition states, several schemes which require polynomial time gates have been proposed to create these states ~\cite{gard2020efficient,wang2009efficient,bergholm2005quantum}. For comparison we also show results for k-particle non-interacting bosons and fermions. Going further, we also show that (k-1)-QW with localized input states distinguish k-CFI graphs. We conclude by proving some results for SRGs where we show that 2-QW with localized input can distinguish SRGs, while 2-QW with input superposition state cannot.

\section{Methods}
\subsection{Quantum Walks and graph isomorphism }
In this work, we focus on continuous quantum walks (CQW) which are implemented via an XY Hamiltonian encode the graphs. The XY Hamiltonian can be written as:
\begin{equation}
    H_{XY}= \sum_{i,j\in E(G)}\sigma_i^x \sigma_j^x + \sigma_i^y \sigma_j^y
\end{equation}
where $E(G)$ is the edge set of graph $G$ and $\sigma_i^x,\sigma_i^y,\sigma_i^z$ are the local Pauli operators.
Beginning with an input state $|\psi_{in}\rangle$, the output state after XY evolution is given by:
\begin{equation}
    |\psi_{out}\rangle = e^{-iH_{XY}\theta}|\psi_{in}\rangle
\end{equation}
The XY evolution is particle conserving, meaning that beginning with an input state with a fixed number of particles, the output state will also have the same number of particles. Hence the evolution happens in a restricted subspace.
For example, the 1-particle subspace is spanned by $|1,0,0..,0\rangle,|0,1,0...,0\rangle,..,|0,0,...,1\rangle$.
The matrix element $\langle i| H_{XY}|j\rangle $ is non-zero only if and only if $(i,j)$ is an edge of $G$.
Therefore, the matrix representation of the Hamiltonian in the z-basis, restricted to the 1-particle manifold, is identical to the adjacency matrix of $G$.\\
This allows us to understand the evolution of the XY Hamiltonian using effective graphs which we call \textit{occupation graphs}.
For a given integer $k\geq 1$, we note $|i_1,i_2,..i_k \rangle$ the k-particle state with particles located on nodes $\{i_1,i_2,..i_k\}$ (forming a basis of the k-particle subspace) and $H^{(k)}_{XY}$ the XY Hamiltonian restricted to this subspace.
For two k-tuples $(i_1,i_2,.i_\alpha..i_k)$ and $(j_1,j_2,.,j_\beta,.,j_k)$, the matrix element $\langle i_1,i_2,..i_k|H^{(k)}_{XY}|j_1,j_2,..j_k \rangle$, is non-zero if and only if the set difference between the two k-tuples has exactly 2 entries, one each from each k-tuple, say $i_\alpha,j_\beta$ such that  $(i_{\alpha}j_{\beta})$ is an edge in $E(G)$.
This corresponds to having a particle on node $j_{\beta}$ of G hopping to node $i_{\alpha}$.\\
%This is because the terms in the XY Hamiltonian are of order 2 and can cause utmost a single particle hop when acting on a particular computational basis state. 
The matrix representation of $H^{(k)}_{XY}$ in this restricted basis defines the adjacency matrix of the k-particle occupation graph $G^{(k)}$.
%This gives rise to the occupation graphs 
In $G^{(k)}$, nodes are labeled by k-tuples and an edge exists between two k-tuples if the corresponding matrix entry for $H^{(k)}$ is non-zero.
For example, a k-tuple $(1,2,3)$ is connected to a k-tuple $(1,3,5)$ in occupation graph $G^{(3)}$ only if $(2,5)\in E(G)$.
It is important to note that, because of the structure of the XY model, any node can have at most 1 particle.
This meas that valid k-tuples are not allowed to contain duplicate indices.
For example, in the case of $G^{(3)}$, nodes like $(1,1,2)$ or $(1,1,1)$ would be forbidden. 
%Thus states with more than 1 particle at a site are forbidden.
This is illustrated in Figure \ref{double_occupation}.
%Hence nodes where $i_\alpha=i_\beta$ for some $\alpha,\beta$ in $(i_1,i_2,..i_\alpha,..,i_\beta,..i_k)$ are isolated or forbidden nodes.

We also note the importance of using different initial states.
For example one can make a choice whether to use the uniform superposition state or a localized input state.
The uniform superposition state for 1-particle would be:
\begin{equation}
   |\mathbf{1}^{(1)}\rangle = \frac{1}{\sqrt{N}}\sum_{i=1}^N|i\rangle.
\end{equation}
More generally, the uniform superposition state for the $k$-particle subspace is given by
\begin{equation}
   |\mathbf{1}^{(k)}\rangle = \frac{1}{\sqrt{\binom{N}{k}}}\sum_{i_1,\ldots,\i_k}|i_1,\ldots,i_k\rangle
\end{equation}

On the other hand, localized input states are states where particles are localized at a single node in the occupation graph $G^{(k)}$.

For a general input state $|\psi_{in}^{(k)}\rangle$ which belongs to the k-particle sub-space, the XY evolution can be written as
\begin{equation}
\begin{split}
    e^{-iH_{XY}\theta}|\psi_{in}^{(k)}\rangle&=(1-i\theta[G^{(k)}] +\frac{i^2\theta^2}{2}[G^{(k)}]^2+..\\
    &+\frac{(-i)^m\theta^m}{m!}[G^{(k)}]^m+..)|\psi_{in} ^{(k)}\rangle
\end{split}
\label{eq:mhop}
\end{equation}\\
where $[G^{(k)}]$ is the adjacency matrix for occupation graph $G^{(k)}$. The $i,j$ entry of $[G^{(k)}]^m$ tells us the number of $m$-hop paths between $k$-tuples $i,j$.\\
The choice of the initial state also affects the distinguishing ability of $H_{XY}$. To see this we consider the problem of trying to distinguish regular graphs (i.e. graphs where all nodes have the same degree) of same degree $d$ and number of nodes $N$ using 1-particle $H_{XY}^{(1)}$. 
For regular graphs the 1-particle uniform superposition state is simply an eigenstate of the 1-particle $H_{XY}^{(1)}$, which is also the adjacency matrix of the graph.
The eigenvalues are the degrees of the nodes. Thus
\begin{equation}
    |\psi_{out1}\rangle = e^{-iH_{XY}^{(1)}(G_1)\theta}|\mathbf{1}^{(1)}\rangle = e^{-i\,  d\,  \theta}|\mathbf{1}^{(1)}\rangle
\end{equation}
and 
\begin{equation}
    |\psi_{out2}\rangle = e^{-iH_{XY}^{(1)}(G_2)}|\mathbf{1}^{(1)}\rangle = e^{-i \, d\,  \theta}|\mathbf{1}^{(1)}\rangle = |\psi_{out1}\rangle
\end{equation}
It is then clear that the $XY$ evolution will not be able to distinguish the 2 graphs, since it will produce the same output state on both graphs.
On the other hand, using localized input states, it is possible to distinguish 2 regular graphs of same degree and size. For example, consider a set of two regular graphs, one which consists of a hexagon and the other which consists of 2 triangles. Both are regular graphs of degree 2 and cannot be distinguished by 1-QW with input state $|\mathbf{1}^{(1)}\rangle$. But using localized input states, $P(|i\rangle\rightarrow |i\rangle)$, where $|i\rangle$ is a state localized at a particular node $i$, will be different for both the graphs since one has a cycle of length 3 and the other does not.

The k-occupation graph $G^{(k)}$ is similar to effective graphs generated during k-WL tests.
These higher-dimensional graphs need storage of order $O(N^{k})$ to store the adjacency matrix and $O(N^{k})$ iterations to reach stable-coloring.
Similarly, if one begins with input superposition states, then we need to measure $O(N^{k})$ quantities corresponding to probabilities to find k-particles at $O(N^{k})$ different locations.
Note that if a QPU platform like neutral atoms is used, then we don't need to explicitly store the higher dimensional graph.
Instead, we have to create a k-particle system and the $XY$ interaction ensures the effective dynamics takes place on $G^{(k)}$.\\
\begin{figure*}[t!]\centering
\includegraphics[scale=0.35]{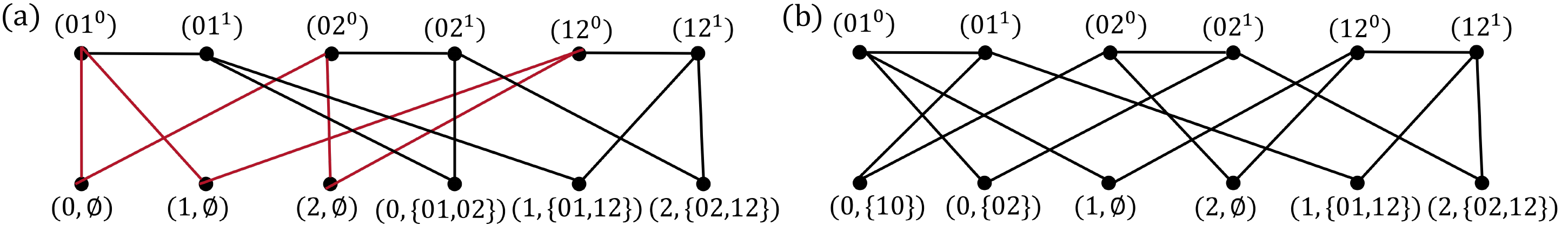}
\caption{Figure shows $k=2$ CFI graph. (a) has a 2-clique of length 3 with nodes $(0,\emptyset),(1,\emptyset),(2,\emptyset)$. We have colored the edges which connect these nodes to each other via paths of length 2.(b) has no such corresponding 2-clique. These graphs are constructed using the original complete graph with node set ${0,1,2}$ and edge set $\left[\{0,1 \},\{0, 2\},\{1, 2\} \right]$ .}
\label{CFI_example}
\end{figure*}
\begin{table*}[t]
\centering
    \begin{tabular}{|c|c|c|c|}
    \hline
       Original graphs &Type I & Type II & Type III \\
    \hline
        &\textcolor{red}{$(0,\emptyset)(2,\emptyset)$} & \textcolor{red}{$(12^0)(02^0)$} & $(01^1)(2,\emptyset)$  \\
     $ P_2$   &\textcolor{red}{$(1,\emptyset)(0,\emptyset)$} & \textcolor{red}{$(01^0)(02^0)$} & \textcolor{red}{$(12^1)(2,\emptyset)$}  \\
          & & \textcolor{red}{$(01^0)(12^0)$} & $(1,\emptyset)(02^1)$  \\
          & &  & \textcolor{red}{$(1,\emptyset)(12^1)$}   \\
           \hline
        &$(1,\emptyset)(0,\{10\})$ & \textcolor{red}{$(12^0)(02^0)$} & $(01^1)(2,\emptyset)$  \\
      $Q_2$ & $(0,\{02\})(2,\emptyset)$ & $(01^0)(02^0)$ & \textcolor{red}{$(12^1)(2,\emptyset)$}  \\
          & & \textcolor{red}{$(01^0)(12^0)$} & $(1,\emptyset)(02^1)$  \\
          & &  & \textcolor{red}{$(1,\emptyset)(12^1)$}   \\
    \hline
    \end{tabular}
    \caption{List of all the allowed 2-hop neighbors of $(1,\emptyset)(2,\emptyset)$ in both $P_2^{(2)}$ and $Q_2^{(2)}$. Type I neighbors correspond to having two consecutive hops of the same particle, returning back to a bottom node. Type II neighbors correspond to having one hop of each particle. Type III nodes correspond to having two consecutive hops of the same particle, ending on a top node. We have colored the pairs which constitutes a 2-clique in the original graphs. We can see that $(1,\emptyset)(2,\emptyset)$ has more allowed 2-hop neighbors that form a 2-neighbors in $P_2^{(2)}$. This fact will lead to less allowed neighbors after two more hops.
    }
    \label{table:2hop_neighbours}
\end{table*}
To understand the origin of the expressiveness of quantum walks, we begin by defining input states which belong to a particular particle sector $|\psi_{in}^{(k)}\rangle$ and expanding it in terms of eigenvectors of that sector.
\begin{equation}
    |\psi_{in}^{(k)}\rangle = \sum _\alpha c_\alpha |\phi_{\alpha}^{(k)}\rangle
\end{equation}
Then the XY evolution leads to:
\begin{equation}
    e^{-iH_{XY}\theta}|\psi_{in}^{(k)}\rangle = \sum_{\alpha}c_\alpha e^{-i\lambda_{\alpha}\theta}|\phi_{\alpha}^{(k)}\rangle
\end{equation}
where $\lambda_{\alpha}$ are the eigenvalues.
Then to calculate the probability of detecting the above output state in a state $|i\rangle$, we obtain:
\begin{equation}
    |\langle i|e^{-iH_{XY}\theta}|\psi_{in}^{(k)}\rangle|^2 = \sum_{\alpha,\beta}c_{\alpha}c_{\beta}^*e^{-i(\lambda_{\alpha}-\lambda_{\beta})\theta}\langle i|\phi_{\alpha}^{(k)}\rangle \langle \phi_{\beta}^{(k)}|i\rangle
\label{eq:prob}
\end{equation}
From \ref{eq:prob}, we see that the expressivity comes from the eigenvectors and eigenvalues of the various subspaces.
If we restrict our input state to a certain particle number, the relevant eigenvectors also belong only to that sector of the Hilbert space. \\
In addition we can also use the notion of limiting probability distribution $p_{\infty}$ as a $\theta$-independent property which has been used in past studies \cite{childs2017lecture}. This is defined as:
\begin{equation}
    p_{\infty}(|\psi_{in}^{(k)}\rangle\to |i\rangle) = \lim_{T\rightarrow \infty}\frac{1}{T}\int_{0}^{T}dt|\langle i|e^{-iH_{XY}t}|\psi_{in}^{(k)}\rangle|^2 
\label{eq:pinfty_1}
\end{equation}
Using the expansion of the Hamiltonian $H_{XY}$ using its eigen-projectors, $H_{XY}=\sum_{\lambda \in Sp(H)}\lambda P^{\lambda}$, where $\lambda$ are the eigenvalues and $P^\lambda$ are the projectors into the corresponding eigenspace, we can write the above equation as:
\begin{equation}
    p_{\infty}(|\psi_{in}^{(k)}\rangle\to |i\rangle)=\sum_{\lambda \in Sp(H_{XY})}|\langle i|P^{\lambda}|\psi_{in}^{(k)}\rangle|^2
\label{eq:pinfty_2}
\end{equation}
where we assume for simplicity that the eigenvalue spectrum of $H_{XY}$ is non-degenerate.

We also define a metric $\Delta$ which we use to numerically distinguish non-Isomorphic graphs. A non-zero value of $\Delta$ means that the algorithm is able to distinguish the two graphs. Suppose we begin with an input state $|\psi_{in}\rangle$, then we prepare a list of probabilities $L_{G_1} = \left\{|\langle i|e^{iH_{XY,G_1}\theta}|\psi_{in}\rangle|^2\right\}_i$ where $i$ spans the k-tuples and $H_{XY,G1}$ is the XY Hamiltonian defined for graph $G_1$. Then we define $\Delta$ as :
\begin{equation}
    \Delta = \sum_i |\text{sort}(L_{G_1})-\text{sort}(L_{G_2})|
\end{equation}
The '$\text{sort}$' operation ensures that $\Delta$ is zero for isomorphic graphs. Our decision to use $\Delta$ over other metrics like Kullback-Leibler (KL) divergences or Jensen-Shannon divergences is based on its simplicity and the fact that it allows us to measure distance between graphs when using localized states as inputs since in this case we obtain a multi-list of probabilities which cannot be interpreted readily using KL and Jensen-Shannon divergences.

\subsection{Introduction to SRGS and relation to WL tests}
In this section we discuss SRGs and their family parameters.
SRGs are a popular set of benchmark graphs used to test the distinguishing ability of various algorithms for graph isomorphism.
The SRGs have very uniform local properties. They are defined by 4 family parameters:
\begin{itemize}
    \item N: number nodes in a graph.
    \item k: degree of all the nodes in the graph.
    \item $\mu$: Number of shared neighbors between adjacent nodes.
    \item $\nu$: Number of shared neighbors between non-adjacent nodes.
\end{itemize}
Various databases now exist with several families of SRGs that have been discovered so far~\footnote{\url{http://www.maths.gla.ac.uk/~es/srgraphs.php}}~\footnote{\url{https://math.ihringer.org/srgs.php}}.
More generally SRGs can be generated as line intersection graphs of generalized quadrangles ~\cite{godsil2015}.   

It has been shown that 3-WL or higher are needed for distinguishing SRGs ~\cite{bodnar2021weisfeiler}. SRGs have been also explored using algorithms based on QW ~\cite{gamble2010,Rudinger2013,godsil2015}.
In particular ~\cite{gamble2010} uses 2-particle quantum walks with localized input states to generate a list of probabilities of length $O(N^4)$ to distinguish SRGs.

For any SRG,
\begin{equation}
A^2  = (k-\nu)I + \nu J + (\nu-\mu)A
\end{equation}
where $J$ is the matrix of all ones and $I$ is the identity matrix. The matrices $A,I,J$ therefore form a closed algebra.
This implies that every power of $A$ can be written in terms of SRG family parameters.
Hence the probabilities due to XY evolution would now be functions of only the family parameters.
And therefore the 1-particle QW cannot distinguish SRGs.

\subsection{Introduction to k-CFI graphs and relation to k-WL tests}
CFI graphs were introduced in ~\cite{cai1992} as graphs to challenge the k-WL tests.
The authors describe the construction of a set of graphs that a certain k-WL test cannot distinguish.
However they can be distinguished by a k+1 WL test.
The set of 2 graphs are distinguished by the presence (or absence) of a twist between a certain set of nodes.\\
A similar construction was defined in ~\cite{morris2020weisfeiler} where the 2 rival k-CFI graphs can be distinguished by the presence (or absence) of a {\it 2-clique} of size $k+1$.

A set of nodes $\gamma$ form a 2-clique if all the nodes in the set are at a distance 2 from each other.
We primarily focus on this CFI constructions in the further discussions as this will facilitate a better understanding of the terminology employed and provide clarity to support our evidence-based arguments.

\subsection{CFI graph construction}

We present here the CFI construction as discussed in ~\cite{morris2020weisfeiler}.
We begin with an undirected graph $K$ which is a complete graph consisting of $k+1$ vertices $(0,1,..,k)$.
Assuming $E(v)$ to be the set of edges incident on vertex $v$, we define the graph $P_k$ as follows:
\begin{enumerate}
    \item The vertex set $V(P_k)$ consists of the following vertices:
    For each vertex $v\in V(K)$, we add a vertex $(v,S)$ for each even subset $S$ of $E(v)$.
    This includes the empty set $\emptyset$. In addition for each edge $e \in E(K)$, we add two vertices $e^1,e^0$.
    \item To create the edge set $E(P_k)$ we first add edges between $e^0,e^1$ for all $e \in E(K)$.
    We add an edge between all nodes $(v,S)$ and $e^1$ if $v \in e$ and $e \in S$.
    Further we add an edge for all nodes $(v,S)$ and $e^0$ if $v \in e$ and $e \not \in S$.
\end{enumerate}
The nodes of the type $(v,S)$ have a degree k. This can be seen using the above rule 2.
We call them \textit{top} nodes. $(v,S)$ is connected to $e^1$ for all $e\in E(v) \land e\in S$. Let us call all the edges in $E(v)$ but not in $S$ as $S'$.
Rule 2 then says that $(v,S)$ is connected to all $e^0$ with $e\in S'$.
But we know that $|S|+|S'|=deg(v)=k$. Hence degree of all nodes $(v,S)$ is k.
We also show that degree of nodes of type $e^0$ and $e^1$ (which we call \textit{bottom} nodes) is the same for $k>1$.
To see this, we note that $e^1$ is connected to all nodes $(v,S),v\in e$ and $S$ is the set of even size edges of $v$ where $e$ occurs.
The set of even sets of edges of $v$ in which $e$ occur is $(1/2)2^{N-1}$.
To see this, we note that number of even partitions of a set of N elements $(1/2)2^N=2^{N-1}$.
The number of even partitions in which a certain element occurs is equal to number of odd partitions of $N-1$ elements which is equal to $(1/2)2^{N-1}$.
Therefore the number of even partitions of a set of size $N$ in which a certain element does not occur is $2^{N-1}-(1/2)2^{N-1}=2^{N-2}$.
Therefore the degree of nodes of type $e^0$ and $e^1$ is the same.% for $k>1$ $P_k$ CFI graphs.\\

To create the rival graph $Q_k$, in step 1, instead of choosing even subsets, we choose odd subsets of node 0.
The total number of vertices are $(k+1)2^{k-1}+2\binom{k+1}{2}$ for both $P_k,Q_k$.
In addition the same arguments as for $P^k$ can be used to prove that nodes of type $(v,S)$ have degree $k$.
Moreover using the argument that the number of times an element occurs in odd subsets of a set of size $N$ is equal to the number of times it does not occur and both are equal to $(1/2)2^{N-1}$, we can use the same argument as for $P^k$ to show that $e^0$ and $e^1$ type nodes have the same degree in $P_k$ and $Q_k$ for $k>1$.

The two graphs can be distinguished by the presence or absence of a 2-clique of size k+1.
The two graphs are non-isomorphic because the graph $P_k$ has a 2-clique of size k+1 formed with nodes of degree k, which is not present in $Q_k$ ~\cite{morris2020weisfeiler}.

We show 2-CFI graphs in Figure \ref{CFI_example}.
These graphs are generated from a complete graph with 3 nodes $(0,1,2)$ and edge set ${01,02,12}$. It can be seen that the nodes $(0,\emptyset),(1,\emptyset),(2,\emptyset)$ in Figure \ref{CFI_example}(a) form a 2-clique of size 3.
There is no corresponding 2-clique of size 3 in Figure \ref{CFI_example}(b).

\begin{table*}[t]
\centering
    \begin{tabular}{|c|c|c|c|c|c|c|}
    \hline
        k&size of graph $G$&size of graph $G^{(k)}$& $\Delta$  & $\Delta(p_{\infty})$ &  $\Delta$ N.I. Bosons & $\Delta$ with N.I. Fermion\\
    \hline
        1 & 4 & 4 & 0.6436 & 0.3667& 0.6436 & 0.6436   \\
    \hline
        2 & 12 & 66 & 0.1923 & 0.0619 & 4.7e-16 & 0.1302  \\
     \hline   
        3 & 28 & 3276 & 0.0084 &0.0028 & 7.96e-15 & 0.0082 \\
    \hline
    \end{tabular}
    \caption{Gap size using k-QW with input superposition states to distinguish k-CFI graphs defined in \cite{morris2020weisfeiler}. We also show results for non-interacting Bosons and Fermions for comparison with input uniform superposition states.}
    \label{table:1}
\end{table*}

\begin{table*}[t]
\centering
    \begin{tabular}{|c|c|c|c|c|c|c|}
    \hline
        k&size of graph $G$&size of graph $G^{(k)}$& $\Delta$ & $\Delta(p_{\infty})$& $\Delta$ with N.I. Bosons & $\Delta$ with N.I. Fermion\\
    \hline
        1 & 6 & 6 & 0.0835 &0.1667 & 0.0835 & 0.0835 \\
    \hline
        2 & 18 & 153 & 0.4994 & 0.2611 & 3.09e-16 & 0.0535  \\
     \hline   
        3 & 40 & 9880 & 0.0018 & 0.0003 & 4.443e-15 & 0.0039 \\
    \hline
    \end{tabular}
    \caption{Gap size using k-QW with input superposition states to distinguish k-CFI graphs defined in \cite{cai1992}. We also show results for non-interacting Bosons and Fermions for comparison with input uniform superposition states.}
    \label{table:2}
\end{table*}

\section{Results}
\subsection{k-CFI graphs can be distinguished using k-QW with input uniform superposition states.}
We first prove the trivial case for k=1 CFI. Suppose we have nodes 0,1 connected in $G$ by an edge $e\equiv 01$.
Then $P_1$ will have the vertices $[(0,\emptyset),(1,\emptyset),01^0,01^1]$ and edges $[\{(0,\emptyset),01^0\},\{01^0,01^1\},\{01^0,(1,\emptyset)\}]$.
The graph $Q_1$ will have the nodes $[(0,{01}),(1,\emptyset),01^0,01^1]$ and edge set $\left[\{(0,{01}),01^1\},\{01^0,01^1\},\{01^0,(1,\emptyset)\}\right]$.
Thus $01^0$ has degree 3 in $P_1$ while there is no nodes in $Q_1$ with degree 3. Thus these graphs are trivially distinguished because of the difference in the degree distribution. 

For $k>1$, proof is defined in two steps:
\begin{enumerate}

\item
We first prove that there are a set of unique k-tuples in graph $P_k$ such that the number of effective children after various number of hops is unique to both $P_k,Q_k$.
\item
We then show that this is enough for k-QW beginning with input superposition states to distinguish the 2 graphs.
\end{enumerate}

We first discuss the idea behind the proof and then describe it more explicitly while illustrating it with an example for $k=2$.
To prove that a k-QW with input uniform superposition states can distinguish between two k-CFI graphs, we consider the various $m$-hop neighbors of k-tuples in these graphs.
In particular we consider the k-tuple $((1,\emptyset),(2,\emptyset),\ldots (k,\emptyset))$.
The entries of this k-tuple along with the node $(0,\emptyset)$ form a size $k+1$ 2-clique in graph $P_k$.
All the nodes in this set have a degree $k$ by construction.
We will show that graph $Q_k$ does not have a $k$-tuple with nodes of degree $k$, such that the number of $m$-hop neighbors is the same as for $((1,\emptyset),(2,\emptyset),\ldots (k,\emptyset))$ for any value of $m$.

Considering hops from $(1,\emptyset)$ and leaving the other vertices untouched, the k-tuple  $((1,\emptyset),(2,\emptyset),\dots(k,\emptyset))$ has 2-hop neighbors $((2,\emptyset),(2,\emptyset),\ldots (k,\emptyset))$, $((3,\emptyset),(2,\emptyset),(3,\emptyset),...(k,\emptyset))$,..,$((k,\emptyset),(2,\emptyset),..(k,\emptyset))$. All these k-tuples contain repeated entries of nodes, which are forbidden in the QW occupation graph. Similar forbidden nodes can be generated by considering 2-hops of the other components of the $k$-tuple. The corresponding k-tuple in graph $Q_k$ with all nodes of degree $k$ which can have the same number of QW-forbidden entries after 2-hops is a k-tuple of nodes which form a 2-clique of size $k$, which is possible in graph $Q_k$. We label this k-tuple as $((1',\emptyset),(2',\emptyset),..(k',\emptyset))$.

The graph $P_k$ will have further $k$ more 2-hop neighbors of the kind $((0,\emptyset),(2,\emptyset),\ldots,(k,\emptyset))$, $((1,\emptyset),(0,\emptyset),\ldots(k,\emptyset))$,\ldots,$((1,\emptyset),(2,\emptyset),\ldots (0,\emptyset))$.
This is because $(0,\emptyset)$ forms a size $k+1$ 2- clique with entries in $((1,\emptyset),(2,\emptyset),\ldots(k,\emptyset))$. 
Each of these $k$ branches will after 2 more hops lead to k-tuples with QW-forbidden nodes.
Similarly the other $k-1$ 2- hop neighbors will lead to QW-forbidden k-tuples after 2 more hops.
No such evolution is possible for the corresponding k-tuple in $Q_k$ $((1',\emptyset),(2',\emptyset),\ldots(k',\emptyset))$, since this would imply that there is a size $k+1$ 2-clique in $Q_k$ which is a contradiction.

To see this, suppose the nodes $((1',\emptyset),(2',\emptyset),\ldots(k'-1,\emptyset))$ have a shared 2-neighbor $(0',\emptyset)$.
At most 1 such node is possible since the degree of the nodes $((1',\emptyset),(2',\emptyset),\ldots(k',\emptyset))$ is $k$.
Then we can only have $k-1$ branches after 2 hops of the kind $(0',\emptyset),(2',\emptyset),\ldots,(k',\emptyset)$, $(1',\emptyset),(0',\emptyset),\ldots,(k',\emptyset)$ up to $(1',\emptyset),(2',\emptyset),\ldots,(0',\emptyset),(k',\emptyset)$.
After 2 more hops, we will still have QW forbidden nodes, but less in quantity because $(0',\emptyset)$ is not a 2-neighbor of $(k',\emptyset)$.
The total number of $m$-hop paths for any k-tuple with all $k$-degree nodes is the same, since the local neighborhood of all the $k$-degree nodes is the same in both graphs. Here $m$-hop paths refer to different paths of length $m$ which a particle can take beginning from a certain node.
However because of QW-forbidden nodes, the number of effective m-hop paths for k-tuple $((1,\emptyset),(2,\emptyset),\ldots(k,\emptyset))$ is not equal to any $k$-tuple in graph $Q_k$.

To see this more explicitly, we illustrate this part of the proof for $k$=2 in Table \ref{table:2hop_neighbours}, in which we define the three types of 2-hop neighbors and list those of $(1,\emptyset),(2,\emptyset)$ in both $P_2^{(2)}$ and $Q_2^{(2)}$.
From the table we see that there are 3 types of tuples after 2 hops which can potentially lead to tuples with repeated entries after 2 more hops.
\begin{enumerate}
    \item
The type I nodes can lead to repeated entry nodes after 2 more hops since $(0,\emptyset),(1,\emptyset),(2,\emptyset)$ form a size 3 2-clique which has not present in $Q_2$.
\item For the type 2 nodes, $(12^0)(02^0)$ and $(01^0)(12^0)$ lead to repeated entry nodes for both $P_k,Q_k$ because it just depends on whether $(1,\emptyset),(2,\emptyset)$ share a common neighbor(forms a 2-clique).
However $(01^0)(02^0)$ leads to a repeated entry nodes only in $P_k$ because $(0,\emptyset),(1,\emptyset),(2,\emptyset)$ forms a size 3 2-clique and no such 3 clique exists in $Q_k$.
\item Type 3 nodes consist of tuples where 2 hops in only 1 entry of the tuple leads to a tuple with a node of type $e^1$. 
Only in the case where $e^1$ is connected to a node which is a common neighbor of $(1,\emptyset),(2,\emptyset)$ we can obtain nodes with repeated entry after 2 more hops.
Thus since $(1,\emptyset),(2,\emptyset)$ share a common neighbor in both $P_k,Q_k$, type 3 nodes will lead to same number of forbidden nodes after 2 more hops.
\end{enumerate}
We thus see that number of forbidden nodes after 2 hops and after 4 hops depends on whether entries in tuples either share a common neighbor or whether they share a common 2 neighbor, $P_2$ leads to more forbidden nodes after 2 and 4 levels than any 2 tuple in $Q_2$.

To generalize this, we first show that $(1,\emptyset),(2,\emptyset)...(k,\emptyset)$ in $P_k$ can be distinguished from any $k$-tuple  which has a smaller 2-clique in the k elements in $Q_k$ in 2 hops.
Suppose we have a general tuple $(a,b,c\ldots)$.
Then forbidden nodes can happen if and only if either say $a,b$ share a common neighbor and there is a 1-hop from each to this common neighbor or if $a$ makes 2 hops and reaches $b$ or vice-versa. Both require that $a,b$ are 2-neighbors.
The number of such possibilities is $n\choose 2$ where $n$ is the number of elements which are 2-neighbors of each other in the k-tuple.
Clearly for a k-tuple this number is maximum only when all $k$ elements are 2 neighbors of each other.

To distinguish two k-tuples in $P_k$ and $Q_k$ in which all $k$ elements are 2-neighbors of each other, we have to go to 4 hops.
Note that only in case of $P_k$ can the k-elements be a part of size k+1 clique.
For 4 hops we consider different length partitions of 4:
\begin{enumerate}[label=\arabic*.]
    \item Size 1 partition : In this case only 1 particle makes 4 hops. Suppose $(a,b,c,..)$ form a k-tuple.
    Then this can happen in the following 2 ways:
    \begin{enumerate}[label=(\roman*)]
        \item Suppose $a,b$ share a common 2 neighbor, $x$, not in the k-tuple.
        Then particle hops from $a$ to $x$ in 2 hops and then to $b$ in 2 more hops.
        \item Suppose $a,b$ share a common neighbor.
        Then particle from $a$ goes to $e^1$ node which is a neighbor of a common neighbor of $a,b$ in 2 hops.
        In 2 more hops it reached $b$.
    \end{enumerate}

\item Size 2 partition: (2+2).
In this case a particle makes 2 hops from one location and another particle makes 2 more hops to reach the same location.
Just like for 1, this can happen in 2 ways:
\begin{enumerate}[label=(\roman*)]
    \item 
Suppose $(a,b,c\ldots)$ form the k-tuple and $a,b$ share a common 2 neighbor outside or inside the k-tuple, $x$.
Then a particle makes 2 hops from $a$ to $x$ and a particle from $b$ also makes 2 hops to reach $x$.
\item $a,b$ share a common neighbor $x$.
A particle from $a$ takes 2 hops to reach $e^1$ which is a neighbor of $x$. Similarly a particle from $b$ reaches $e^1$ in 2 hops.
\end{enumerate}
\item Size 2 partition : (3+1).
This can also happen in 2 ways:
\begin{enumerate}[label=(\roman*)]
    \item 
Suppose $(a,b,c..)$ form the k-tuple. Suppose $a,b$ have common 2 -neighbor $x$.
Then in 3 hops a particle can reach from $a$ to $x$ and then to a common neighbor between $b$ and $x$.
\item Suppose $a,b$ share a common neighbor $x$.
Then a particle takes 3 hops to go from $a$ to $x$ to $e^1$ and then back to $x$.
The second particle takes 1 hop from $b$ to $x$.
\end{enumerate}
\item Size 3 partition: (2+1+1).
2 hops take a particle to either a bottom node or a node of type $e^1$.
On the other hand 1 hop takes a particle to the top node of type $e^0$.
Thus a forbidden node is possible only of the particles taking 1-hops have a common neighbor of type $e^0$.
\item Size 4 partition:(1+1+1+1):
We have repeated nodes if any of set of 4 nodes have common neighbors.
\end{enumerate}
From the above cases we can see that the number of possible forbidden nodes depends on the set of nodes being 2- neighbors of each other or having a set of common 2 neighbors outside the k-tuple. The first condition holds equal for both $P_k$ and $Q_k$ when we choose a k-tuple which also form a size $k$ 2-clique. In case of $Q_k$ however, at most $k-1$ nodes from this $k$-tuple can have shared common 2 neighbor outside the $k$-tuple. Thus we can have at most $(k-1)\choose 2$ ways to have forbidden nodes where bottom nodes outside the $k$-tuple are involved. On the other hand for $P_k$ since a size $k+1$ 2-clique is possible, this number is $k\choose 2$. Thus $P_k$ with $k$-tuple which is a part of size $k+1$ 2-clique will have more forbidden neighbors after 4 hops than any k-tuple of $Q_k$ which also forms a size $k$-clique.

To see that this is enough for QW with input superposition state to distinguish the two graphs, let us suppose that two k-tuples in $P_k$ indexed with 'i' and $Q_k$ indexed with 'j' have same effective neighbors in their respective k-occupation graphs $P_k^{(k)}$ and $Q_k^{(k)}$ up to $m$-1 hops and differ in the $m^{th}$ hop. Therefore, $([P_k^{(k)}]^m|\mathbf{1}^{(k)}\rangle)_i\ne([Q_k^{(k)}]^m|\mathbf{1}^{(k)}\rangle)_j$, where $[P_k^{(k)}]$ and $[Q_k^{(k)}]$ are adjacency matrices corresponding to occupation graphs $P_k^{(k)}$ and $Q_k^{(k)}$ respectively. Since there is no k-tuple in $Q_k$ which has the same evolution as $((1,\emptyset),(2,\emptyset),..(k,\emptyset))$, if $((1,\emptyset),(2,\emptyset),..(k,\emptyset))$ is indexed 'i' in $P_k^{(k)}$, there will be some $m$ for which $([P_k^{(k)}]^m|\mathbf{1}^{(k)}\rangle)_i\ne([Q_k^{(k)}]^m|\mathbf{1}^{(k)}\rangle)_j$ for any $k$-tuple in $Q_k$. Since (from \ref{eq:mhop}), the evolution under $H_{XY}$ is a sum of m-hop terms, for a suitable range of values of $\theta$, evolution of $|1^{(k)}\rangle$, for the two graphs would lead to output state vectors not related by a permutation.  \\
Tables \ref{table:1},\ref{table:2} shows the empirical results for k-QW for k-CFI graphs. We see that the occupation graph sizes increase rapidly and we are able to simulate up to k=3 CFI graphs and observe that they are indeed distinguished by k-QW. For comparison, we also provide $\Delta$ for non-interacting Bosons and Fermions. The definitions of bosonic or fermionic basis states (particle-on-vertices basis) and their evolution are based on definitions used in ~\cite{Rudinger2012}. From the results, we notice that we obtain similar values of $\Delta$ for evolution using $H_{XY}$ and non-interacting fermions, while obtaining very low values for non-interacting bosons. This could be because just like for occupation graphs for $H_{XY}$, nodes which correspond to multiple particles at the same site are forbidden for non-interacting fermions, while they are allowed for non-interacting bosons. We also notice the values of $\Delta$ seem to be decreasing with increasing $k$. Although, it is difficult to conclude from data up to only k=3, a possible reason could be that proportion of set of k-tuples which are different in the two graphs is lesser for higher k. We also have calculated values of $\Delta$ using $p_{\infty}$ using definition \ref{eq:pinfty_1} as shown in the tables.

\subsection{k-CFI can be distinguished by (k-1) QW with localized input}
We first show that 1-particle QW can distinguish k=2 CFI graphs for localized input excitations.  For this we consider the circuit probability for nodes which form a 2-clique in graph $P_2$, i.e. $P(i\rightarrow i)$. In particular we consider paths where except the first and the last nodes, none of the nodes are repeated. Since this node ‘i’ is part of a 2-triangle (triangle, but where the edges represent a distance 2) because it is a part of a size 3 2-clique, and since there is no corresponding 2-triangle in $Q_2$, a particle can go from $i\rightarrow i$ in 6-hops in $P_2$, while there is no such path available in $Q_2$. Hence at most 6 single particle hops are enough to distinguish k=2 CFI graphs.   

\begin{figure}[h!]\centering
\includegraphics[scale=0.4]{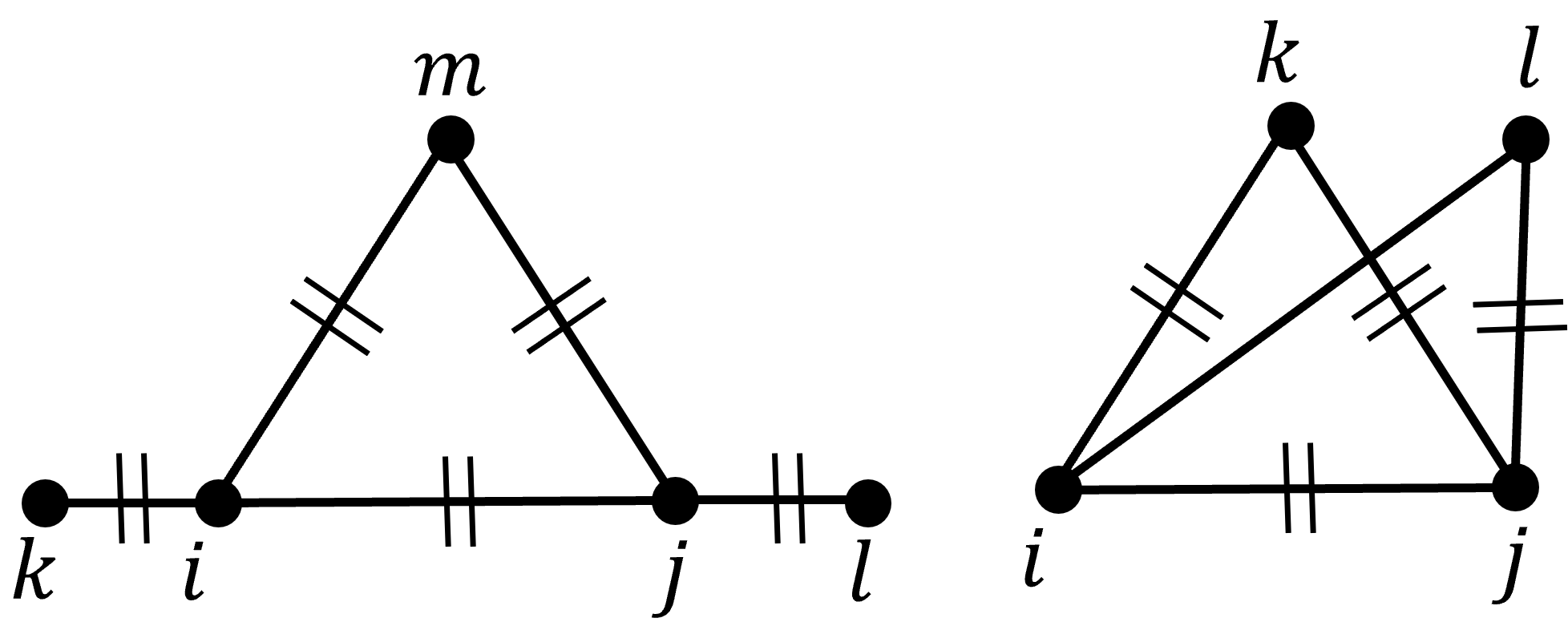}
\caption{Figure shows how number of common neighbors will affect the outcome of 2 walks. For example since the number of 2-triangles (where each edge is of path length 2)  is different in the 2 figures, the number of ways particles go from $ij\rightarrow ij$ will be different. The 2 dashed lines on the edges represent a distance 2 between corresponding nodes.}
\label{Common_neighbors}
\end{figure}

For the general k, we first illustrate the proof using the case for 2-QW for 3-CFI graphs, one of which has a size 4 2-clique. We prove that there is no tuple like $ij$ (which are a part of a size 4 2-clique in $P_3$ as shown in Figure \ref{2CFI_2clique}) in the rival graph $Q_3$ with the following properties:
\begin{figure}[h!]\centering
\includegraphics[scale=0.4]{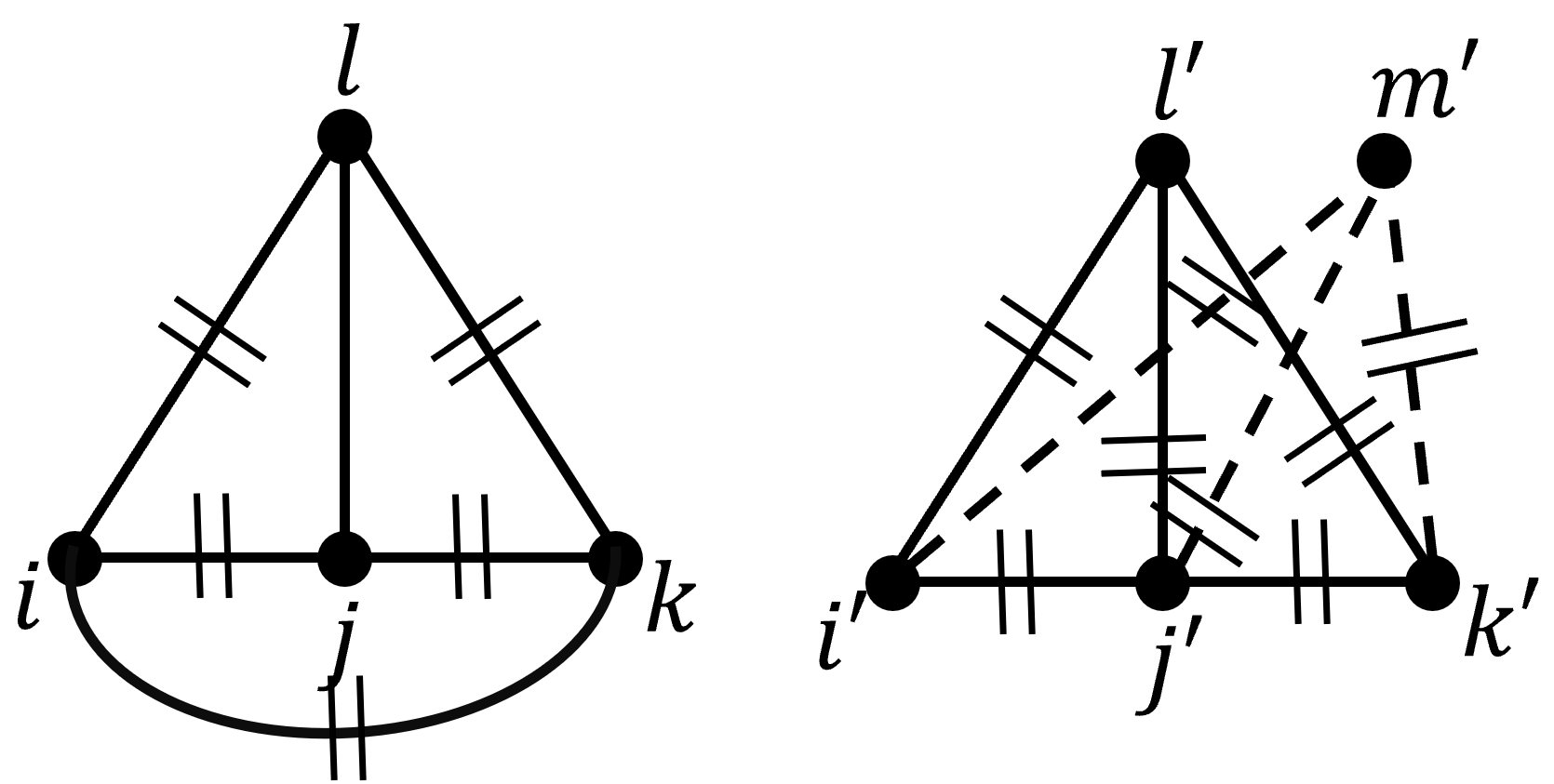}
\caption{Figure shows a size 4 2-clique $i,j,k,l$ in the 3-CFI graph. A similar attempt to create it using $i,j,k,l,m$ leads to node $j$ having a degree 4 which leads to a contradiction. The 2 dashed lines on the edges represent a distance 2 between the corresponding nodes}
\label{2CFI_2clique}
\end{figure}
\begin{enumerate}
\item Same degree: This can be checked simply by checking the degree of $i$ and $j$ in the adjacency matrix. For a fixed degree of $i$ and $j$ the degree of $ij$ is determined by its isomorphism type. By isomorphism type, we mean whether $i,j$ is an edge or an empty graph.
Using the 2-QW, the first hop itself is good enough to reveal the isomorphism type. In addition we also require the nodes to be 2-neighbors of each other. The 2-QW can detect this because after 2 hops the evolution will lead to 2 forbidden nodes $ii$ and $jj$. Thus effective number of nodes reached will be less if $i,j$ are 2-neighbors. This argument can be similarly extended for $k-1$ tuples and $k-1$ QW. 
\item Same number of shared 2-neighbors: The number of shared 2-neighbors for the same degree and isomorphism type of $ij$ determines its circuit probability. Figure \ref{Common_neighbors} illustrates this. $P(ij\rightarrow ij)$ after a certain number of hops is different depending on number of closed paths and will be detected by the 2-QW. This is because number of 2-triangles (triangles but where each edge of the triangle represents a path of length 2) will be different.
\item Has a 2-neighbor of type $jk$ of the same degree and isomorphism type and the same number of shared neighbors as $ij$. The 2-neighbor of type $jk$ can be found easily since it differs from $ij$ in only one entry and nodes in $jk$ form a 2-clique. Hence after 2 hops the number of forbidden nodes will be maximum for $jk$. Moreover, trying to enforce a node of the type $jk$ in the rival graph leads to a contradiction as we can show. Any other kind of $jk$ can be detected as discussed above.
\end{enumerate}

To prove the contradiction, consider the size 4 2-clique in Figure \ref{2CFI_2clique}. We choose this clique with nodes having degree 3 as can be seen for $i,j,k,l$. The number of shared 2-neighbors between $i,j$ is 2 ($l,k$). Similarly $j,k$ have shared 2-neighbors $i,l$. Figure shows what happens when we try to enforce this condition without a size 4 2-clique. $i',j'$ should have same shared 2-neighbors as $i,j$. Suppose we call them $l',m'$. Now we need $j',k'$ to also have 2 shared 2-neighbors. But we also need $j'k'$ to have the same isomorphism type as $jk$. This induces a contradiction as this requires $j'$ to have a degree 4.\\
Therefore suppose one begins with a particle at node $ij$. Then we check the degree of the nodes$(i,j)$ and then its isomorphism type which can be checked in 1 hop. Then one looks at number of nodes after 2 hops to check whether $ij$ forms a 2 clique. The number of shared neighbors can be found from the circuit probability. Thereafter we look at its neighbors of type $jk$. It should differ in 1 entry from $ij$, should be a 2-clique and should have same number of common neighbors as $ij$. Both these properties can be found similar to $ij$ by looking at degrees of the nodes, isomorphism type and circuit probabilities. 
Such a neighbor of $i'j'$ does not exist for the rival graph.
\begin{figure}[h!]\centering
\includegraphics[scale=0.45]{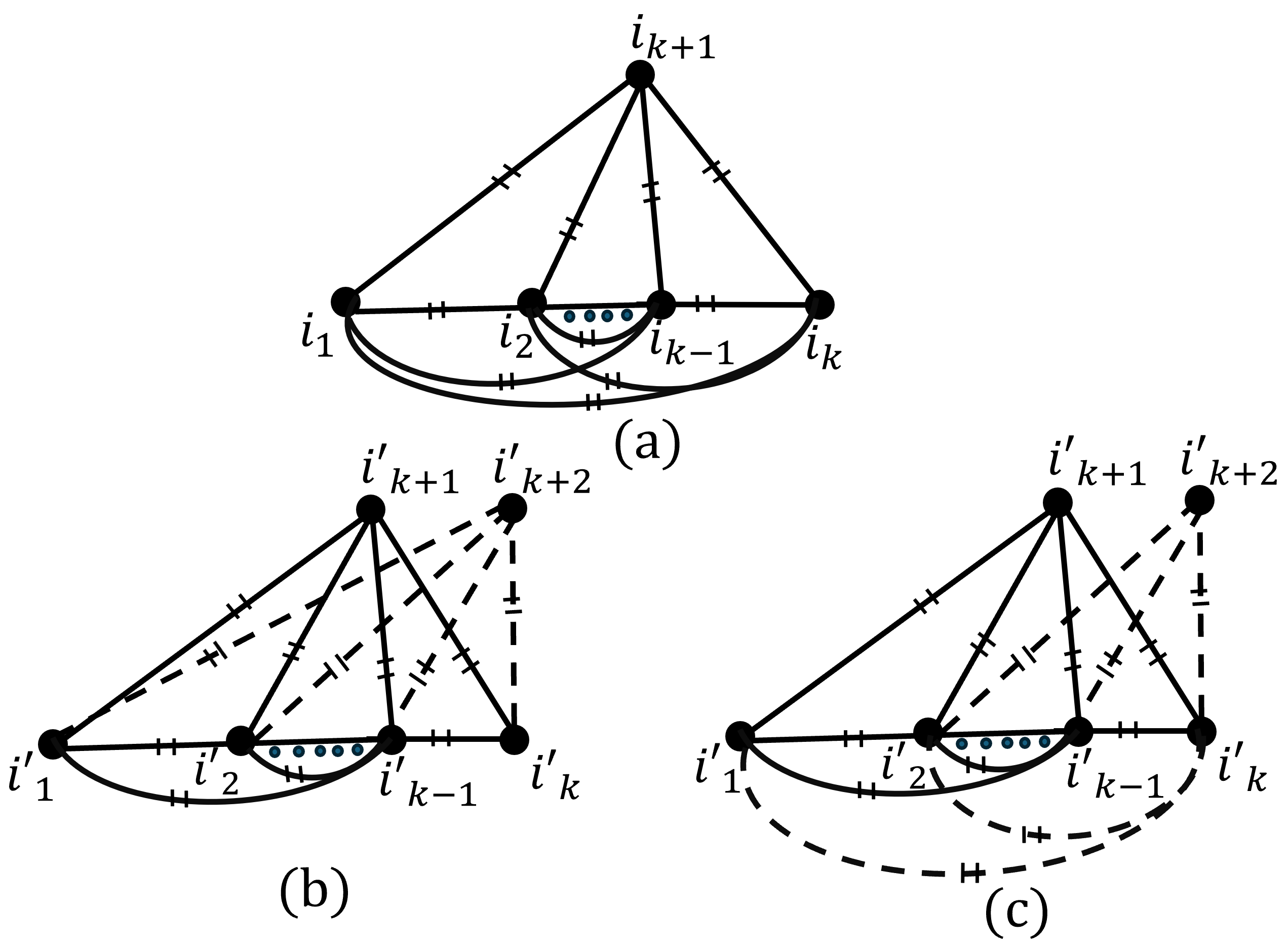}
\caption{Figure (a) shows the size $k+1$ 2-clique in the $k$-CFI graph. (b) and (c) shows the two conditions which can happen when we try to create a tuple $i'_1,i'_2,...,i'_{k-1}$ with same properties as $i_1,i_2,...,i_{k-1}$ in the rival graph. The 2 dashed lines on the edges represent a distance 2 between corresponding nodes.}
\label{general_k}
\end{figure}

The above contradiction can be seen also for the general case of k-1 QW for k-CFI graphs, by considering 2-cliques of size k+1. Suppose the nodes are labelled $i_1,i_2,..,i_{k+1}$. Then nodes $i_1,i_2,...,i_{k-1}$ share 2 common 2-neighbors $i_k,i_{k+1}$. Similarly $i_2,i_3,..,i_{k}$ share 2 common 2-neighbors $i_1,i_{k+1}$. The degrees of each of all these nodes is $k$. Suppose we now try to create a node $i'_1,i'_2,...,i'_{k-1}$ in the rival graph with the analogous properties to points 1,2 above. Since it has to have the same isomorphism type and number of shared neighbors as $i_1,i_2,..i_{k-1}$, we can have 2 cases (illustrated in Figure \ref{general_k}):
\begin{enumerate}
\item
1. The first option is $i'_{k+1},i'_{k+2}$. We also further require $i'_2,i'_3,...i'_{k}$ to have 2 share neighbors. We can choose again $i'_{k+1},i'_{k+2}$. However, this would require degrees of nodes $i'_2,...,i'_{k-1}$ to be k+1 which is a contradiction.
\item The second option is $i'_{k},i'_{k+1}$. Also then $i'_2,i'_3,..i'_{k}$ would have 2 common 2-neighbors $i_{k+1},i_{k+2}$. However this would also need the degree of $i'_2,i'_3,..,i_{k-1}$ to be $k+1$ which is a contradiction.
\end{enumerate}
In Tables \ref{table:3} and \ref{table:4}, we show the empirical results for k=2 and k=3 CFI graphs using 1-QW and 2-QW respectively with localized input states and see that we are able to distinguish them using the $H_{XY}$ interaction. On the other hand, while for k=2 CFI graphs we obtain large values for $\Delta$ when using 1-particle bosonic or fermionic particles, a much lower value of $\Delta$ is obtained for k=3 CFI graphs. This also highlights the fact that the traversal over k-occupation graphs is very different than traversal by a set of non-interacting particles. We also have calculated values of $\Delta$ using $p_{\infty}$ using definition \ref{eq:pinfty_2}, since we obtained large errors using definition $\ref{eq:pinfty_1}$. We choose different values of T and stop till convergence within a certain error-bar as shown in the tables.      
\begin{table*}[t]
\centering
    \begin{tabular}{|c|c|c|c|c|c|c|}
    \hline
        k&size of graph $G$&size of graph $G^{(k)}$& $\Delta$ & $\Delta(p_{\infty})$& $\Delta$ with N.I. Bosons & $\Delta$ with N.I. Fermion\\
    \hline
        2 & 12 & 12 & 6.4823 & 1.95 $\pm$ 0.05 & 6.4823 & 6.4823 \\
    \hline
        3 & 28 & 378 & 9.5073 & 6.47 $\pm$ 0.04 & 1.13e-13 & 1.25e-13  \\
     
    \hline
    \end{tabular}
    \caption{Gap size using k-1 QW with input superposition states to distinguish k-CFI graphs defined in ~\cite{morris2020weisfeiler}. We also show results for non-interacting Bosons and Fermions for comparison with input uniform superposition states.}
    \label{table:3}
\end{table*}

\begin{table*}[t]
\centering
    \begin{tabular}{|c|c|c|c|c|c|c|}
    \hline
        k&size of graph $G$&size of graph $G^{(k)}$& $\Delta$ &$\Delta(p_{\infty})$ &$\Delta$ with N.I. Bosons & $\Delta$ with N.I. Fermion\\
    \hline
        2 & 18 & 18 & 4.8682 & 16.15 $\pm$ 0.05 & 4.8682 & 4.8682 \\
    \hline
        3 & 40 & 780 & 3.5894 & 4.88$\pm$0.04 & 5.446e-13 & 5.789e-13  \\
     
    \hline
    \end{tabular}
    \caption{Gap size using k-1 QW with input superposition states to distinguish k-CFI graphs defined in ~\cite{cai1992}. We also show results for non-interacting Bosons and Fermions for comparison with input uniform superposition states.}
    \label{table:4}
\end{table*}

\subsection{2-QW can distinguish SRGs with localized input}
\begin{figure}[h!]\centering
\includegraphics[scale=0.5]{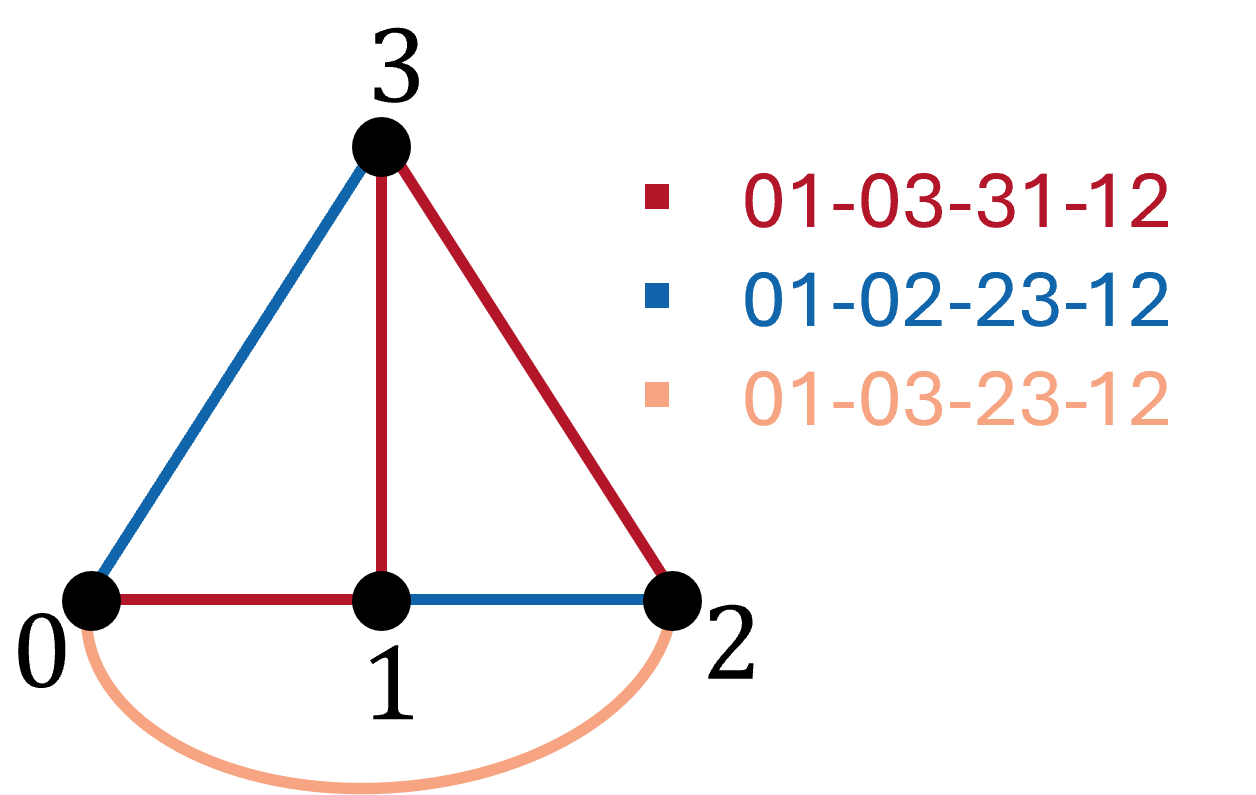}
\caption{Figure shows a 4-clique. This can also be interpreted as node $3$ being a shared neighbor for $0,1,2$. We also show few ways in which we can go from $01\rightarrow 12$ via 3 hops where each of the edges participates in some 3-hop path. Different colors show the edges involved in a certain 3-hop path. Some edges are repeated for multiple paths. Note that only paths where no edge is repeated are shown.}
\label{four_clique}
\end{figure}

In this section we show that 2-QW counts quantities which are not only dependent on the SRG family parameters and hence can distinguish SRGs. For this we show the example shown in Figure \ref{four_clique}. This example shows a 4- clique with 3 being a shared neighbor of $0,1,2$. It can be seen that the number of paths of length 3 between $01\rightarrow 12$ depends on all the edges in the clique. A presence of an additional shared neighbor would increase the number of paths of length 3 between $01\rightarrow 12$. However, the number of shared neighbors between three nodes, is not a SRG family parameter. We know that $[G^{(2)}]^3_{01,12}$, which is 3rd power of the 2-particle occupation graph calculates the number of paths of length 3 between $01,12$. We have seen that this quantity is not dependent on the SRG family parameter and hence a 2-QW can distinguish SRGs. \\
Coincidentally, the presence or absence of a 4-clique in SRGs of 16 nodes using $[G^{(2)}]^3$ is one of the quantities which distinguishes the 2 graphs. 

\subsection{2-QW cannot distinguish SRGs with input uniform superposition}
Each node in graph $uv\in G^{(2)}$ has a degree based on whether it is an edge or a non-edge in the original graph. Suppose $uv$ is a edge, then the number of common neighbors between $u,v$ in $G$ is the family parameter $\mu$. Suppose the common neighbors are $(1,2...,\mu)$, then in $G^{(2)}$, $uv$ is adjacent to $u1,u2,,u\mu$ and $1v,2v,..\mu v$. Therefore the common $\mu$ neighbors contribute to a degree $2\mu$. Moreover, all these neighbors are of edge type. Further, $uv$ is also adjacent to $(d-1-\mu)$ neighbors of the type $uv_1,uv_2,..uv_{d-1-\mu}$ where $v_1,v_2,..v_{d-1-\mu} \in \mathcal{N}_v$ are neighbors of $v$. These neighbors are of non-edge type. Similarly there are neighbors of type $u_1v,u_2v,..u_{d-1-\mu}v$. Therefore $uv$ has $2(d-1-\mu)$ neighbors of non-edge type. Therefore the degree of $uv$ is $2(d-1)$. Similarly it can be shown that $uv$ of non-edge type has degree $2d$ with $2\nu$ neighbors of edge type and $2(d-\nu)$ neighbors of non-edge type.\\    
Therefore the number of 1-hop neighbors for any node $i$ in $G^{(2)}$ and the type of neighbors is given by SRG family parameter. Since the number of 1-hop neighbors for any node $j\in \mathcal{N}_i$ are in-turn given by the family parameters, it can be seen that for any $m$, the number of $m$-hop neighbors of any node in $G^{(2)}$ is given by the family parameters. Since any quantity $([A^{(2)}]^m|\mathbf{1}^{(2)}\rangle)_i$ counts the number of $m$-hop neighbors of the $i^{th}$ node in $G^{(2)}$, these values are given by the SRG family parameters. Hence two SRGs of the same family will have equal number of entries of a given value for any $m$-hop and hence cannot be distinguished by a 2-QW with input superposition state.

\section{Conclusions and Outlook}
We have examined k-QW with different input states and we observed that their distinguishing ability depends on their initial states. We have also studied their ability to distinguish hard graphs like k-CFI and SRGs. In particular, we saw that k-QW with input superposition states distinguish k-CFI graphs. It is also an interesting future problem of interest to see how the value of $\Delta$ scales with $k$. Our results strongly suggest that k-QW have similar distinguishing abilities and complexities as classical k-WL tests. Similar to the k-WL tests which have time-complexity of $O(N^k)$, the number of quantities we measure for $k$-QW with input superposition states goes as $O(N^k)$ while for $k-1$QW with input localized states as $O(N^{2k-2})$. However it is also important to note that unlike for k-WL tests which require $O(N^k)$ storage to store an effective graph, k-QW using hardware platforms like neutral atoms can implement dynamics on an occupation graph $G^{(k)}$ of size $O(N^k)$ simply by creating $k$ excitations in a N-atom system. Although experimentally challenging, this could provide an interesting direction for a quantum advantage on graph-based practical problems of interest.

\begin{acknowledgements}
The authors would like to thank Yann Bouchereau, Igor Sokolov, Mehdi Djellabi, Slimane Thabet and Constantin Dalyac for stimulating and helpful discussions.

\end{acknowledgements}

\bibliography{ref}

\begin{thebibliography}{38}%
\makeatletter
\providecommand \@ifxundefined [1]{%
 \@ifx{#1\undefined}
}%
\providecommand \@ifnum [1]{%
 \ifnum #1\expandafter \@firstoftwo
 \else \expandafter \@secondoftwo
 \fi
}%
\providecommand \@ifx [1]{%
 \ifx #1\expandafter \@firstoftwo
 \else \expandafter \@secondoftwo
 \fi
}%
\providecommand \natexlab [1]{#1}%
\providecommand \enquote  [1]{``#1''}%
\providecommand \bibnamefont  [1]{#1}%
\providecommand \bibfnamefont [1]{#1}%
\providecommand \citenamefont [1]{#1}%
\providecommand \href@noop [0]{\@secondoftwo}%
\providecommand \href [0]{\begingroup \@sanitize@url \@href}%
\providecommand \@href[1]{\@@startlink{#1}\@@href}%
\providecommand \@@href[1]{\endgroup#1\@@endlink}%
\providecommand \@sanitize@url [0]{\catcode `\\12\catcode `\$12\catcode `\&12\catcode `\#12\catcode `\^12\catcode `\_12\catcode `\%12\relax}%
\providecommand \@@startlink[1]{}%
\providecommand \@@endlink[0]{}%
\providecommand \url  [0]{\begingroup\@sanitize@url \@url }%
\providecommand \@url [1]{\endgroup\@href {#1}{\urlprefix }}%
\providecommand \urlprefix  [0]{URL }%
\providecommand \Eprint [0]{\href }%
\providecommand \doibase [0]{https://doi.org/}%
\providecommand \selectlanguage [0]{\@gobble}%
\providecommand \bibinfo  [0]{\@secondoftwo}%
\providecommand \bibfield  [0]{\@secondoftwo}%
\providecommand \translation [1]{[#1]}%
\providecommand \BibitemOpen [0]{}%
\providecommand \bibitemStop [0]{}%
\providecommand \bibitemNoStop [0]{.\EOS\space}%
\providecommand \EOS [0]{\spacefactor3000\relax}%
\providecommand \BibitemShut  [1]{\csname bibitem#1\endcsname}%
\let\auto@bib@innerbib\@empty
%</preamble>
\bibitem [{\citenamefont {Childs}(2009)}]{childs2009}%
  \BibitemOpen
  \bibfield  {author} {\bibinfo {author} {\bibfnamefont {A.~M.}\ \bibnamefont {Childs}},\ }\href {https://doi.org/10.1103/PhysRevLett.102.180501} {\bibfield  {journal} {\bibinfo  {journal} {Physical Review Letters}\ }\textbf {\bibinfo {volume} {102}},\ \bibinfo {pages} {180501} (\bibinfo {year} {2009})}\BibitemShut {NoStop}%
\bibitem [{\citenamefont {Childs}\ and\ \citenamefont {Goldstone}(2004)}]{childs2004}%
  \BibitemOpen
  \bibfield  {author} {\bibinfo {author} {\bibfnamefont {A.~M.}\ \bibnamefont {Childs}}\ and\ \bibinfo {author} {\bibfnamefont {J.}~\bibnamefont {Goldstone}},\ }\href {https://doi.org/10.1103/PhysRevA.70.022314} {\bibfield  {journal} {\bibinfo  {journal} {Physical Review A}\ }\textbf {\bibinfo {volume} {70}},\ \bibinfo {pages} {022314} (\bibinfo {year} {2004})}\BibitemShut {NoStop}%
\bibitem [{\citenamefont {Childs}\ and\ \citenamefont {Berry}(2012)}]{childs2012}%
  \BibitemOpen
  \bibfield  {author} {\bibinfo {author} {\bibfnamefont {A.~M.}\ \bibnamefont {Childs}}\ and\ \bibinfo {author} {\bibfnamefont {D.~W.}\ \bibnamefont {Berry}},\ }\href {https://doi.org/10.26421/QIC12.1-2-4} {\bibfield  {journal} {\bibinfo  {journal} {Quantum Information and Computation}\ }\textbf {\bibinfo {volume} {12}},\ \bibinfo {pages} {29} (\bibinfo {year} {2012})}\BibitemShut {NoStop}%
\bibitem [{\citenamefont {Mohseni}\ \emph {et~al.}(2008)\citenamefont {Mohseni}, \citenamefont {Rebentrost}, \citenamefont {Lloyd},\ and\ \citenamefont {Aspuru-Guzik}}]{mohseni2008}%
  \BibitemOpen
  \bibfield  {author} {\bibinfo {author} {\bibfnamefont {M.}~\bibnamefont {Mohseni}}, \bibinfo {author} {\bibfnamefont {P.}~\bibnamefont {Rebentrost}}, \bibinfo {author} {\bibfnamefont {S.}~\bibnamefont {Lloyd}},\ and\ \bibinfo {author} {\bibfnamefont {A.}~\bibnamefont {Aspuru-Guzik}},\ }\href {https://doi.org/10.1063/1.3002335} {\bibfield  {journal} {\bibinfo  {journal} {The Journal of Chemical Physics}\ }\textbf {\bibinfo {volume} {129}},\ \bibinfo {pages} {174106} (\bibinfo {year} {2008})}\BibitemShut {NoStop}%
\bibitem [{\citenamefont {Childs}\ \emph {et~al.}(2003)\citenamefont {Childs}, \citenamefont {Cleve}, \citenamefont {Deotto}, \citenamefont {Farhi}, \citenamefont {Gutmann},\ and\ \citenamefont {Spielman}}]{childs2003}%
  \BibitemOpen
  \bibfield  {author} {\bibinfo {author} {\bibfnamefont {A.~M.}\ \bibnamefont {Childs}}, \bibinfo {author} {\bibfnamefont {R.}~\bibnamefont {Cleve}}, \bibinfo {author} {\bibfnamefont {E.}~\bibnamefont {Deotto}}, \bibinfo {author} {\bibfnamefont {E.}~\bibnamefont {Farhi}}, \bibinfo {author} {\bibfnamefont {S.}~\bibnamefont {Gutmann}},\ and\ \bibinfo {author} {\bibfnamefont {D.~A.}\ \bibnamefont {Spielman}},\ }in\ \href {https://doi.org/10.1145/780542.780552} {\emph {\bibinfo {booktitle} {Proceedings of the thirty-fifth annual ACM symposium on Theory of computing}}}\ (\bibinfo  {publisher} {ACM},\ \bibinfo {address} {San Diego CA USA},\ \bibinfo {year} {2003})\ pp.\ \bibinfo {pages} {59--68}\BibitemShut {NoStop}%
\bibitem [{\citenamefont {Krovi}\ and\ \citenamefont {Brun}(2006)}]{krovi2006}%
  \BibitemOpen
  \bibfield  {author} {\bibinfo {author} {\bibfnamefont {H.}~\bibnamefont {Krovi}}\ and\ \bibinfo {author} {\bibfnamefont {T.~A.}\ \bibnamefont {Brun}},\ }\href {https://doi.org/10.1103/PhysRevA.73.032341} {\bibfield  {journal} {\bibinfo  {journal} {Physical Review A}\ }\textbf {\bibinfo {volume} {73}},\ \bibinfo {pages} {032341} (\bibinfo {year} {2006})}\BibitemShut {NoStop}%
\bibitem [{\citenamefont {Balasubramanian}\ \emph {et~al.}(2023)\citenamefont {Balasubramanian}, \citenamefont {Li},\ and\ \citenamefont {Harrow}}]{balasubramanian2023exponential}%
  \BibitemOpen
  \bibfield  {author} {\bibinfo {author} {\bibfnamefont {S.}~\bibnamefont {Balasubramanian}}, \bibinfo {author} {\bibfnamefont {T.}~\bibnamefont {Li}},\ and\ \bibinfo {author} {\bibfnamefont {A.}~\bibnamefont {Harrow}},\ }\href@noop {} {\bibinfo {title} {Exponential speedups for quantum walks in random hierarchical graphs}} (\bibinfo {year} {2023}),\ \Eprint {https://arxiv.org/abs/2307.15062} {arXiv:2307.15062 [quant-ph]} \BibitemShut {NoStop}%
\bibitem [{\citenamefont {Schulz}\ \emph {et~al.}(2024)\citenamefont {Schulz}, \citenamefont {Willsch},\ and\ \citenamefont {Michielsen}}]{schulz2024}%
  \BibitemOpen
  \bibfield  {author} {\bibinfo {author} {\bibfnamefont {S.}~\bibnamefont {Schulz}}, \bibinfo {author} {\bibfnamefont {D.}~\bibnamefont {Willsch}},\ and\ \bibinfo {author} {\bibfnamefont {K.}~\bibnamefont {Michielsen}},\ }\href {https://doi.org/10.1103/PhysRevResearch.6.013312} {\bibfield  {journal} {\bibinfo  {journal} {Physical Review Research}\ }\textbf {\bibinfo {volume} {6}},\ \bibinfo {pages} {013312} (\bibinfo {year} {2024})}\BibitemShut {NoStop}%
\bibitem [{\citenamefont {Gamble}\ \emph {et~al.}(2010)\citenamefont {Gamble}, \citenamefont {Friesen}, \citenamefont {Zhou}, \citenamefont {Joynt},\ and\ \citenamefont {Coppersmith}}]{gamble2010}%
  \BibitemOpen
  \bibfield  {author} {\bibinfo {author} {\bibfnamefont {J.~K.}\ \bibnamefont {Gamble}}, \bibinfo {author} {\bibfnamefont {M.}~\bibnamefont {Friesen}}, \bibinfo {author} {\bibfnamefont {D.}~\bibnamefont {Zhou}}, \bibinfo {author} {\bibfnamefont {R.}~\bibnamefont {Joynt}},\ and\ \bibinfo {author} {\bibfnamefont {S.~N.}\ \bibnamefont {Coppersmith}},\ }\href {https://doi.org/10.1103/PhysRevA.81.052313} {\bibfield  {journal} {\bibinfo  {journal} {Physical Review A}\ }\textbf {\bibinfo {volume} {81}},\ \bibinfo {pages} {052313} (\bibinfo {year} {2010})}\BibitemShut {NoStop}%
\bibitem [{\citenamefont {Rudinger}\ \emph {et~al.}(2013)\citenamefont {Rudinger}, \citenamefont {Gamble}, \citenamefont {Bach}, \citenamefont {Friesen}, \citenamefont {Joynt},\ and\ \citenamefont {Coppersmith}}]{Rudinger2013}%
  \BibitemOpen
  \bibfield  {author} {\bibinfo {author} {\bibfnamefont {K.}~\bibnamefont {Rudinger}}, \bibinfo {author} {\bibfnamefont {J.~K.}\ \bibnamefont {Gamble}}, \bibinfo {author} {\bibfnamefont {E.}~\bibnamefont {Bach}}, \bibinfo {author} {\bibfnamefont {M.}~\bibnamefont {Friesen}}, \bibinfo {author} {\bibfnamefont {R.}~\bibnamefont {Joynt}},\ and\ \bibinfo {author} {\bibfnamefont {S.~N.}\ \bibnamefont {Coppersmith}},\ }\href {https://doi.org/doi:10.1166/jctn.2013.3105} {\bibfield  {journal} {\bibinfo  {journal} {Journal of Computational and Theoretical Nanoscience}\ }\textbf {\bibinfo {volume} {10}},\ \bibinfo {pages} {1653} (\bibinfo {year} {2013})}\BibitemShut {NoStop}%
\bibitem [{\citenamefont {Conte}\ \emph {et~al.}(2004)\citenamefont {Conte}, \citenamefont {Foggia}, \citenamefont {Sansone},\ and\ \citenamefont {Vento}}]{conte2004}%
  \BibitemOpen
  \bibfield  {author} {\bibinfo {author} {\bibfnamefont {D.}~\bibnamefont {Conte}}, \bibinfo {author} {\bibfnamefont {P.}~\bibnamefont {Foggia}}, \bibinfo {author} {\bibfnamefont {C.}~\bibnamefont {Sansone}},\ and\ \bibinfo {author} {\bibfnamefont {M.}~\bibnamefont {Vento}},\ }\href {https://doi.org/10.1142/S0218001404003228} {\bibfield  {journal} {\bibinfo  {journal} {International Journal of Pattern Recognition and Artificial Intelligence}\ }\textbf {\bibinfo {volume} {18}},\ \bibinfo {pages} {265} (\bibinfo {year} {2004})}\BibitemShut {NoStop}%
\bibitem [{\citenamefont {Merkys}\ \emph {et~al.}(2023)\citenamefont {Merkys}, \citenamefont {Vaitkus}, \citenamefont {Grybauskas}, \citenamefont {Konovalovas}, \citenamefont {Quirós},\ and\ \citenamefont {Gražulis}}]{merkys2023}%
  \BibitemOpen
  \bibfield  {author} {\bibinfo {author} {\bibfnamefont {A.}~\bibnamefont {Merkys}}, \bibinfo {author} {\bibfnamefont {A.}~\bibnamefont {Vaitkus}}, \bibinfo {author} {\bibfnamefont {A.}~\bibnamefont {Grybauskas}}, \bibinfo {author} {\bibfnamefont {A.}~\bibnamefont {Konovalovas}}, \bibinfo {author} {\bibfnamefont {M.}~\bibnamefont {Quirós}},\ and\ \bibinfo {author} {\bibfnamefont {S.}~\bibnamefont {Gražulis}},\ }\href {https://doi.org/10.1186/s13321-023-00692-1} {\bibfield  {journal} {\bibinfo  {journal} {Journal of Cheminformatics}\ }\textbf {\bibinfo {volume} {15}},\ \bibinfo {pages} {25} (\bibinfo {year} {2023})}\BibitemShut {NoStop}%
\bibitem [{\citenamefont {Abiad}\ \emph {et~al.}(2020)\citenamefont {Abiad}, \citenamefont {Grigoriev},\ and\ \citenamefont {Niemzok}}]{ABIAD2020}%
  \BibitemOpen
  \bibfield  {author} {\bibinfo {author} {\bibfnamefont {A.}~\bibnamefont {Abiad}}, \bibinfo {author} {\bibfnamefont {A.}~\bibnamefont {Grigoriev}},\ and\ \bibinfo {author} {\bibfnamefont {S.}~\bibnamefont {Niemzok}},\ }\href {https://doi.org/https://doi.org/10.1016/j.cie.2020.106715} {\bibfield  {journal} {\bibinfo  {journal} {Computers and Industrial Engineering}\ }\textbf {\bibinfo {volume} {148}},\ \bibinfo {pages} {106715} (\bibinfo {year} {2020})}\BibitemShut {NoStop}%
\bibitem [{\citenamefont {Babai}(2016)}]{babai2016}%
  \BibitemOpen
  \bibfield  {author} {\bibinfo {author} {\bibfnamefont {L.}~\bibnamefont {Babai}},\ }\href@noop {} {\bibinfo {title} {Graph isomorphism in quasipolynomial time}} (\bibinfo {year} {2016}),\ \Eprint {https://arxiv.org/abs/1512.03547} {arXiv:1512.03547 [cs.DS]} \BibitemShut {NoStop}%
\bibitem [{\citenamefont {Weisfeiler}\ and\ \citenamefont {Leman}(1968)}]{weisfeiler1968reduction}%
  \BibitemOpen
  \bibfield  {author} {\bibinfo {author} {\bibfnamefont {B.}~\bibnamefont {Weisfeiler}}\ and\ \bibinfo {author} {\bibfnamefont {A.}~\bibnamefont {Leman}},\ }\href@noop {} {\bibfield  {journal} {\bibinfo  {journal} {nti, Series}\ }\textbf {\bibinfo {volume} {2}},\ \bibinfo {pages} {12} (\bibinfo {year} {1968})}\BibitemShut {NoStop}%
\bibitem [{\citenamefont {Grohe}\ and\ \citenamefont {Otto}(2015)}]{grohe2015pebble}%
  \BibitemOpen
  \bibfield  {author} {\bibinfo {author} {\bibfnamefont {M.}~\bibnamefont {Grohe}}\ and\ \bibinfo {author} {\bibfnamefont {M.}~\bibnamefont {Otto}},\ }\href@noop {} {\bibinfo {title} {Pebble games and linear equations}} (\bibinfo {year} {2015}),\ \Eprint {https://arxiv.org/abs/1204.1990} {arXiv:1204.1990 [cs.LO]} \BibitemShut {NoStop}%
\bibitem [{\citenamefont {Grohe}(2017)}]{grohe2017descriptive}%
  \BibitemOpen
  \bibfield  {author} {\bibinfo {author} {\bibfnamefont {M.}~\bibnamefont {Grohe}},\ }\href@noop {} {\emph {\bibinfo {title} {Descriptive complexity, canonisation, and definable graph structure theory}}},\ Vol.~\bibinfo {volume} {47}\ (\bibinfo  {publisher} {Cambridge University Press},\ \bibinfo {year} {2017})\BibitemShut {NoStop}%
\bibitem [{\citenamefont {Du}\ \emph {et~al.}(2003)\citenamefont {Du}, \citenamefont {Li}, \citenamefont {Xu}, \citenamefont {Shi}, \citenamefont {Wu}, \citenamefont {Zhou},\ and\ \citenamefont {Han}}]{du2003}%
  \BibitemOpen
  \bibfield  {author} {\bibinfo {author} {\bibfnamefont {J.}~\bibnamefont {Du}}, \bibinfo {author} {\bibfnamefont {H.}~\bibnamefont {Li}}, \bibinfo {author} {\bibfnamefont {X.}~\bibnamefont {Xu}}, \bibinfo {author} {\bibfnamefont {M.}~\bibnamefont {Shi}}, \bibinfo {author} {\bibfnamefont {J.}~\bibnamefont {Wu}}, \bibinfo {author} {\bibfnamefont {X.}~\bibnamefont {Zhou}},\ and\ \bibinfo {author} {\bibfnamefont {R.}~\bibnamefont {Han}},\ }\href {https://doi.org/10.1103/PhysRevA.67.042316} {\bibfield  {journal} {\bibinfo  {journal} {Physical Review A}\ }\textbf {\bibinfo {volume} {67}},\ \bibinfo {pages} {042316} (\bibinfo {year} {2003})}\BibitemShut {NoStop}%
\bibitem [{\citenamefont {Karski}\ \emph {et~al.}(2009)\citenamefont {Karski}, \citenamefont {Förster}, \citenamefont {Choi}, \citenamefont {Steffen}, \citenamefont {Alt}, \citenamefont {Meschede},\ and\ \citenamefont {Widera}}]{karski2009}%
  \BibitemOpen
  \bibfield  {author} {\bibinfo {author} {\bibfnamefont {M.}~\bibnamefont {Karski}}, \bibinfo {author} {\bibfnamefont {L.}~\bibnamefont {Förster}}, \bibinfo {author} {\bibfnamefont {J.-M.}\ \bibnamefont {Choi}}, \bibinfo {author} {\bibfnamefont {A.}~\bibnamefont {Steffen}}, \bibinfo {author} {\bibfnamefont {W.}~\bibnamefont {Alt}}, \bibinfo {author} {\bibfnamefont {D.}~\bibnamefont {Meschede}},\ and\ \bibinfo {author} {\bibfnamefont {A.}~\bibnamefont {Widera}},\ }\href {https://doi.org/10.1126/science.1174436} {\bibfield  {journal} {\bibinfo  {journal} {Science}\ }\textbf {\bibinfo {volume} {325}},\ \bibinfo {pages} {174} (\bibinfo {year} {2009})}\BibitemShut {NoStop}%
\bibitem [{\citenamefont {Schmitz}\ \emph {et~al.}(2009)\citenamefont {Schmitz}, \citenamefont {Matjeschk}, \citenamefont {Schneider}, \citenamefont {Glueckert}, \citenamefont {Enderlein}, \citenamefont {Huber},\ and\ \citenamefont {Schaetz}}]{schmitz2009}%
  \BibitemOpen
  \bibfield  {author} {\bibinfo {author} {\bibfnamefont {H.}~\bibnamefont {Schmitz}}, \bibinfo {author} {\bibfnamefont {R.}~\bibnamefont {Matjeschk}}, \bibinfo {author} {\bibfnamefont {C.}~\bibnamefont {Schneider}}, \bibinfo {author} {\bibfnamefont {J.}~\bibnamefont {Glueckert}}, \bibinfo {author} {\bibfnamefont {M.}~\bibnamefont {Enderlein}}, \bibinfo {author} {\bibfnamefont {T.}~\bibnamefont {Huber}},\ and\ \bibinfo {author} {\bibfnamefont {T.}~\bibnamefont {Schaetz}},\ }\href {https://doi.org/10.1103/PhysRevLett.103.090504} {\bibfield  {journal} {\bibinfo  {journal} {Physical Review Letters}\ }\textbf {\bibinfo {volume} {103}},\ \bibinfo {pages} {090504} (\bibinfo {year} {2009})}\BibitemShut {NoStop}%
\bibitem [{\citenamefont {Peruzzo}\ \emph {et~al.}(2010)\citenamefont {Peruzzo}, \citenamefont {Lobino}, \citenamefont {Matthews}, \citenamefont {Matsuda}, \citenamefont {Politi}, \citenamefont {Poulios}, \citenamefont {Zhou}, \citenamefont {Lahini}, \citenamefont {Ismail}, \citenamefont {Wörhoff}, \citenamefont {Bromberg}, \citenamefont {Silberberg}, \citenamefont {Thompson},\ and\ \citenamefont {OBrien}}]{peruzzo2010}%
  \BibitemOpen
  \bibfield  {author} {\bibinfo {author} {\bibfnamefont {A.}~\bibnamefont {Peruzzo}}, \bibinfo {author} {\bibfnamefont {M.}~\bibnamefont {Lobino}}, \bibinfo {author} {\bibfnamefont {J.~C.~F.}\ \bibnamefont {Matthews}}, \bibinfo {author} {\bibfnamefont {N.}~\bibnamefont {Matsuda}}, \bibinfo {author} {\bibfnamefont {A.}~\bibnamefont {Politi}}, \bibinfo {author} {\bibfnamefont {K.}~\bibnamefont {Poulios}}, \bibinfo {author} {\bibfnamefont {X.-Q.}\ \bibnamefont {Zhou}}, \bibinfo {author} {\bibfnamefont {Y.}~\bibnamefont {Lahini}}, \bibinfo {author} {\bibfnamefont {N.}~\bibnamefont {Ismail}}, \bibinfo {author} {\bibfnamefont {K.}~\bibnamefont {Wörhoff}}, \bibinfo {author} {\bibfnamefont {Y.}~\bibnamefont {Bromberg}}, \bibinfo {author} {\bibfnamefont {Y.}~\bibnamefont {Silberberg}}, \bibinfo {author} {\bibfnamefont {M.~G.}\ \bibnamefont {Thompson}},\ and\ \bibinfo {author} {\bibfnamefont {J.~L.}\ \bibnamefont {OBrien}},\ }\href {https://doi.org/10.1126/science.1193515} {\bibfield  {journal} {\bibinfo  {journal}
  {Science}\ }\textbf {\bibinfo {volume} {329}},\ \bibinfo {pages} {1500} (\bibinfo {year} {2010})}\BibitemShut {NoStop}%
\bibitem [{\citenamefont {Qiang}\ \emph {et~al.}(2021)\citenamefont {Qiang}, \citenamefont {Wang}, \citenamefont {Xue}, \citenamefont {Ge}, \citenamefont {Chen}, \citenamefont {Liu}, \citenamefont {Huang}, \citenamefont {Fu}, \citenamefont {Xu}, \citenamefont {Yi}, \citenamefont {Xu}, \citenamefont {Deng}, \citenamefont {Wang}, \citenamefont {Meinecke}, \citenamefont {Matthews}, \citenamefont {Cai}, \citenamefont {Yang},\ and\ \citenamefont {Wu}}]{qiang2021}%
  \BibitemOpen
  \bibfield  {author} {\bibinfo {author} {\bibfnamefont {X.}~\bibnamefont {Qiang}}, \bibinfo {author} {\bibfnamefont {Y.}~\bibnamefont {Wang}}, \bibinfo {author} {\bibfnamefont {S.}~\bibnamefont {Xue}}, \bibinfo {author} {\bibfnamefont {R.}~\bibnamefont {Ge}}, \bibinfo {author} {\bibfnamefont {L.}~\bibnamefont {Chen}}, \bibinfo {author} {\bibfnamefont {Y.}~\bibnamefont {Liu}}, \bibinfo {author} {\bibfnamefont {A.}~\bibnamefont {Huang}}, \bibinfo {author} {\bibfnamefont {X.}~\bibnamefont {Fu}}, \bibinfo {author} {\bibfnamefont {P.}~\bibnamefont {Xu}}, \bibinfo {author} {\bibfnamefont {T.}~\bibnamefont {Yi}}, \bibinfo {author} {\bibfnamefont {F.}~\bibnamefont {Xu}}, \bibinfo {author} {\bibfnamefont {M.}~\bibnamefont {Deng}}, \bibinfo {author} {\bibfnamefont {J.~B.}\ \bibnamefont {Wang}}, \bibinfo {author} {\bibfnamefont {J.~D.~A.}\ \bibnamefont {Meinecke}}, \bibinfo {author} {\bibfnamefont {J.~C.~F.}\ \bibnamefont {Matthews}}, \bibinfo {author} {\bibfnamefont {X.}~\bibnamefont {Cai}}, \bibinfo {author}
  {\bibfnamefont {X.}~\bibnamefont {Yang}},\ and\ \bibinfo {author} {\bibfnamefont {J.}~\bibnamefont {Wu}},\ }\href {https://doi.org/10.1126/sciadv.abb8375} {\bibfield  {journal} {\bibinfo  {journal} {Science Advances}\ }\textbf {\bibinfo {volume} {7}},\ \bibinfo {pages} {eabb8375} (\bibinfo {year} {2021})}\BibitemShut {NoStop}%
\bibitem [{\citenamefont {Browaeys}\ and\ \citenamefont {Lahaye}(2020)}]{browaeys2020}%
  \BibitemOpen
  \bibfield  {author} {\bibinfo {author} {\bibfnamefont {A.}~\bibnamefont {Browaeys}}\ and\ \bibinfo {author} {\bibfnamefont {T.}~\bibnamefont {Lahaye}},\ }\href {https://doi.org/10.1038/s41567-019-0733-z} {\bibfield  {journal} {\bibinfo  {journal} {Nature Physics}\ }\textbf {\bibinfo {volume} {16}},\ \bibinfo {pages} {132} (\bibinfo {year} {2020})}\BibitemShut {NoStop}%
\bibitem [{\citenamefont {Struck}\ \emph {et~al.}(2013)\citenamefont {Struck}, \citenamefont {Weinberg}, \citenamefont {Ölschläger}, \citenamefont {Windpassinger}, \citenamefont {Simonet}, \citenamefont {Sengstock}, \citenamefont {Höppner}, \citenamefont {Hauke}, \citenamefont {Eckardt}, \citenamefont {Lewenstein},\ and\ \citenamefont {Mathey}}]{struck2013}%
  \BibitemOpen
  \bibfield  {author} {\bibinfo {author} {\bibfnamefont {J.}~\bibnamefont {Struck}}, \bibinfo {author} {\bibfnamefont {M.}~\bibnamefont {Weinberg}}, \bibinfo {author} {\bibfnamefont {C.}~\bibnamefont {Ölschläger}}, \bibinfo {author} {\bibfnamefont {P.}~\bibnamefont {Windpassinger}}, \bibinfo {author} {\bibfnamefont {J.}~\bibnamefont {Simonet}}, \bibinfo {author} {\bibfnamefont {K.}~\bibnamefont {Sengstock}}, \bibinfo {author} {\bibfnamefont {R.}~\bibnamefont {Höppner}}, \bibinfo {author} {\bibfnamefont {P.}~\bibnamefont {Hauke}}, \bibinfo {author} {\bibfnamefont {A.}~\bibnamefont {Eckardt}}, \bibinfo {author} {\bibfnamefont {M.}~\bibnamefont {Lewenstein}},\ and\ \bibinfo {author} {\bibfnamefont {L.}~\bibnamefont {Mathey}},\ }\href {https://doi.org/10.1038/nphys2750} {\bibfield  {journal} {\bibinfo  {journal} {Nature Physics}\ }\textbf {\bibinfo {volume} {9}},\ \bibinfo {pages} {738} (\bibinfo {year} {2013})}\BibitemShut {NoStop}%
\bibitem [{\citenamefont {Barredo}\ \emph {et~al.}(2015)\citenamefont {Barredo}, \citenamefont {Labuhn}, \citenamefont {Ravets}, \citenamefont {Lahaye}, \citenamefont {Browaeys},\ and\ \citenamefont {Adams}}]{barredo2015}%
  \BibitemOpen
  \bibfield  {author} {\bibinfo {author} {\bibfnamefont {D.}~\bibnamefont {Barredo}}, \bibinfo {author} {\bibfnamefont {H.}~\bibnamefont {Labuhn}}, \bibinfo {author} {\bibfnamefont {S.}~\bibnamefont {Ravets}}, \bibinfo {author} {\bibfnamefont {T.}~\bibnamefont {Lahaye}}, \bibinfo {author} {\bibfnamefont {A.}~\bibnamefont {Browaeys}},\ and\ \bibinfo {author} {\bibfnamefont {C.~S.}\ \bibnamefont {Adams}},\ }\href {https://doi.org/10.1103/PhysRevLett.114.113002} {\bibfield  {journal} {\bibinfo  {journal} {Physical Review Letters}\ }\textbf {\bibinfo {volume} {114}},\ \bibinfo {pages} {113002} (\bibinfo {year} {2015})}\BibitemShut {NoStop}%
\bibitem [{\citenamefont {De~Léséleuc}\ \emph {et~al.}(2019)\citenamefont {De~Léséleuc}, \citenamefont {Lienhard}, \citenamefont {Scholl}, \citenamefont {Barredo}, \citenamefont {Weber}, \citenamefont {Lang}, \citenamefont {Büchler}, \citenamefont {Lahaye},\ and\ \citenamefont {Browaeys}}]{de_leseleuc2019}%
  \BibitemOpen
  \bibfield  {author} {\bibinfo {author} {\bibfnamefont {S.}~\bibnamefont {De~Léséleuc}}, \bibinfo {author} {\bibfnamefont {V.}~\bibnamefont {Lienhard}}, \bibinfo {author} {\bibfnamefont {P.}~\bibnamefont {Scholl}}, \bibinfo {author} {\bibfnamefont {D.}~\bibnamefont {Barredo}}, \bibinfo {author} {\bibfnamefont {S.}~\bibnamefont {Weber}}, \bibinfo {author} {\bibfnamefont {N.}~\bibnamefont {Lang}}, \bibinfo {author} {\bibfnamefont {H.~P.}\ \bibnamefont {Büchler}}, \bibinfo {author} {\bibfnamefont {T.}~\bibnamefont {Lahaye}},\ and\ \bibinfo {author} {\bibfnamefont {A.}~\bibnamefont {Browaeys}},\ }\href {https://doi.org/10.1126/science.aav9105} {\bibfield  {journal} {\bibinfo  {journal} {Science}\ }\textbf {\bibinfo {volume} {365}},\ \bibinfo {pages} {775} (\bibinfo {year} {2019})}\BibitemShut {NoStop}%
\bibitem [{\citenamefont {Dalyac}\ \emph {et~al.}(2024)\citenamefont {Dalyac}, \citenamefont {Leclerc}, \citenamefont {Vignoli}, \citenamefont {Djellabi}, \citenamefont {da~Silva~Coelho}, \citenamefont {Ximenez}, \citenamefont {Dareau}, \citenamefont {Dreon}, \citenamefont {Elfving}, \citenamefont {Signoles}, \citenamefont {Henry},\ and\ \citenamefont {Henriet}}]{dalyac2024graph}%
  \BibitemOpen
  \bibfield  {author} {\bibinfo {author} {\bibfnamefont {C.}~\bibnamefont {Dalyac}}, \bibinfo {author} {\bibfnamefont {L.}~\bibnamefont {Leclerc}}, \bibinfo {author} {\bibfnamefont {L.}~\bibnamefont {Vignoli}}, \bibinfo {author} {\bibfnamefont {M.}~\bibnamefont {Djellabi}}, \bibinfo {author} {\bibfnamefont {W.}~\bibnamefont {da~Silva~Coelho}}, \bibinfo {author} {\bibfnamefont {B.}~\bibnamefont {Ximenez}}, \bibinfo {author} {\bibfnamefont {A.}~\bibnamefont {Dareau}}, \bibinfo {author} {\bibfnamefont {D.}~\bibnamefont {Dreon}}, \bibinfo {author} {\bibfnamefont {V.~E.}\ \bibnamefont {Elfving}}, \bibinfo {author} {\bibfnamefont {A.}~\bibnamefont {Signoles}}, \bibinfo {author} {\bibfnamefont {L.-P.}\ \bibnamefont {Henry}},\ and\ \bibinfo {author} {\bibfnamefont {L.}~\bibnamefont {Henriet}},\ }\href@noop {} {\bibinfo {title} {Graph algorithms with neutral atom quantum processors}} (\bibinfo {year} {2024}),\ \Eprint {https://arxiv.org/abs/2403.11931} {arXiv:2403.11931 [quant-ph]} \BibitemShut {NoStop}%
\bibitem [{\citenamefont {Cai}\ \emph {et~al.}(1992)\citenamefont {Cai}, \citenamefont {F{\"u}rer},\ and\ \citenamefont {Immerman}}]{cai1992}%
  \BibitemOpen
  \bibfield  {author} {\bibinfo {author} {\bibfnamefont {J.-Y.}\ \bibnamefont {Cai}}, \bibinfo {author} {\bibfnamefont {M.}~\bibnamefont {F{\"u}rer}},\ and\ \bibinfo {author} {\bibfnamefont {N.}~\bibnamefont {Immerman}},\ }\href@noop {} {\bibfield  {journal} {\bibinfo  {journal} {Combinatorica}\ }\textbf {\bibinfo {volume} {12}},\ \bibinfo {pages} {389} (\bibinfo {year} {1992})}\BibitemShut {NoStop}%
\bibitem [{\citenamefont {Morris}\ \emph {et~al.}(2020)\citenamefont {Morris}, \citenamefont {Rattan},\ and\ \citenamefont {Mutzel}}]{morris2020weisfeiler}%
  \BibitemOpen
  \bibfield  {author} {\bibinfo {author} {\bibfnamefont {C.}~\bibnamefont {Morris}}, \bibinfo {author} {\bibfnamefont {G.}~\bibnamefont {Rattan}},\ and\ \bibinfo {author} {\bibfnamefont {P.}~\bibnamefont {Mutzel}},\ }\href@noop {} {\bibfield  {journal} {\bibinfo  {journal} {Advances in Neural Information Processing Systems}\ }\textbf {\bibinfo {volume} {33}},\ \bibinfo {pages} {21824} (\bibinfo {year} {2020})}\BibitemShut {NoStop}%
\bibitem [{\citenamefont {Gard}\ \emph {et~al.}(2020)\citenamefont {Gard}, \citenamefont {Zhu}, \citenamefont {Barron}, \citenamefont {Mayhall}, \citenamefont {Economou},\ and\ \citenamefont {Barnes}}]{gard2020efficient}%
  \BibitemOpen
  \bibfield  {author} {\bibinfo {author} {\bibfnamefont {B.~T.}\ \bibnamefont {Gard}}, \bibinfo {author} {\bibfnamefont {L.}~\bibnamefont {Zhu}}, \bibinfo {author} {\bibfnamefont {G.~S.}\ \bibnamefont {Barron}}, \bibinfo {author} {\bibfnamefont {N.~J.}\ \bibnamefont {Mayhall}}, \bibinfo {author} {\bibfnamefont {S.~E.}\ \bibnamefont {Economou}},\ and\ \bibinfo {author} {\bibfnamefont {E.}~\bibnamefont {Barnes}},\ }\href@noop {} {\bibfield  {journal} {\bibinfo  {journal} {npj Quantum Information}\ }\textbf {\bibinfo {volume} {6}},\ \bibinfo {pages} {10} (\bibinfo {year} {2020})}\BibitemShut {NoStop}%
\bibitem [{\citenamefont {Wang}\ \emph {et~al.}(2009)\citenamefont {Wang}, \citenamefont {Ashhab},\ and\ \citenamefont {Nori}}]{wang2009efficient}%
  \BibitemOpen
  \bibfield  {author} {\bibinfo {author} {\bibfnamefont {H.}~\bibnamefont {Wang}}, \bibinfo {author} {\bibfnamefont {S.}~\bibnamefont {Ashhab}},\ and\ \bibinfo {author} {\bibfnamefont {F.}~\bibnamefont {Nori}},\ }\href@noop {} {\bibfield  {journal} {\bibinfo  {journal} {Physical Review A}\ }\textbf {\bibinfo {volume} {79}},\ \bibinfo {pages} {042335} (\bibinfo {year} {2009})}\BibitemShut {NoStop}%
\bibitem [{\citenamefont {Bergholm}\ \emph {et~al.}(2005)\citenamefont {Bergholm}, \citenamefont {Vartiainen}, \citenamefont {M{\"o}tt{\"o}nen},\ and\ \citenamefont {Salomaa}}]{bergholm2005quantum}%
  \BibitemOpen
  \bibfield  {author} {\bibinfo {author} {\bibfnamefont {V.}~\bibnamefont {Bergholm}}, \bibinfo {author} {\bibfnamefont {J.~J.}\ \bibnamefont {Vartiainen}}, \bibinfo {author} {\bibfnamefont {M.}~\bibnamefont {M{\"o}tt{\"o}nen}},\ and\ \bibinfo {author} {\bibfnamefont {M.~M.}\ \bibnamefont {Salomaa}},\ }\href@noop {} {\bibfield  {journal} {\bibinfo  {journal} {Physical Review A}\ }\textbf {\bibinfo {volume} {71}},\ \bibinfo {pages} {052330} (\bibinfo {year} {2005})}\BibitemShut {NoStop}%
\bibitem [{\citenamefont {Childs}(2017)}]{childs2017lecture}%
  \BibitemOpen
  \bibfield  {author} {\bibinfo {author} {\bibfnamefont {A.~M.}\ \bibnamefont {Childs}},\ }\href@noop {} {\bibfield  {journal} {\bibinfo  {journal} {Lecture notes at University of Maryland}\ } (\bibinfo {year} {2017})}\BibitemShut {NoStop}%
\bibitem [{Note1()}]{Note1}%
  \BibitemOpen
  \bibinfo {note} {\protect \url {http://www.maths.gla.ac.uk/~es/srgraphs.php}}\BibitemShut {NoStop}%
\bibitem [{Note2()}]{Note2}%
  \BibitemOpen
  \bibinfo {note} {\protect \url {https://math.ihringer.org/srgs.php}}\BibitemShut {NoStop}%
\bibitem [{\citenamefont {Godsil}\ \emph {et~al.}(2015)\citenamefont {Godsil}, \citenamefont {Guo},\ and\ \citenamefont {Myklebust}}]{godsil2015}%
  \BibitemOpen
  \bibfield  {author} {\bibinfo {author} {\bibfnamefont {C.}~\bibnamefont {Godsil}}, \bibinfo {author} {\bibfnamefont {K.}~\bibnamefont {Guo}},\ and\ \bibinfo {author} {\bibfnamefont {T.~G.~J.}\ \bibnamefont {Myklebust}},\ }\href@noop {} {\bibinfo {title} {Quantum walks on generalized quadrangles}} (\bibinfo {year} {2015}),\ \Eprint {https://arxiv.org/abs/1511.01962} {arXiv:1511.01962 [math.CO]} \BibitemShut {NoStop}%
\bibitem [{\citenamefont {Bodnar}\ \emph {et~al.}(2021)\citenamefont {Bodnar}, \citenamefont {Frasca}, \citenamefont {Wang}, \citenamefont {Otter}, \citenamefont {Montufar}, \citenamefont {Lio},\ and\ \citenamefont {Bronstein}}]{bodnar2021weisfeiler}%
  \BibitemOpen
  \bibfield  {author} {\bibinfo {author} {\bibfnamefont {C.}~\bibnamefont {Bodnar}}, \bibinfo {author} {\bibfnamefont {F.}~\bibnamefont {Frasca}}, \bibinfo {author} {\bibfnamefont {Y.}~\bibnamefont {Wang}}, \bibinfo {author} {\bibfnamefont {N.}~\bibnamefont {Otter}}, \bibinfo {author} {\bibfnamefont {G.~F.}\ \bibnamefont {Montufar}}, \bibinfo {author} {\bibfnamefont {P.}~\bibnamefont {Lio}},\ and\ \bibinfo {author} {\bibfnamefont {M.}~\bibnamefont {Bronstein}},\ }in\ \href@noop {} {\emph {\bibinfo {booktitle} {International Conference on Machine Learning}}}\ (\bibinfo {year} {2021})\ pp.\ \bibinfo {pages} {1026--1037}\BibitemShut {NoStop}%
\bibitem [{\citenamefont {Rudinger}\ \emph {et~al.}(2012)\citenamefont {Rudinger}, \citenamefont {Gamble}, \citenamefont {Wellons}, \citenamefont {Bach}, \citenamefont {Friesen}, \citenamefont {Joynt},\ and\ \citenamefont {Coppersmith}}]{Rudinger2012}%
  \BibitemOpen
  \bibfield  {author} {\bibinfo {author} {\bibfnamefont {K.}~\bibnamefont {Rudinger}}, \bibinfo {author} {\bibfnamefont {J.~K.}\ \bibnamefont {Gamble}}, \bibinfo {author} {\bibfnamefont {M.}~\bibnamefont {Wellons}}, \bibinfo {author} {\bibfnamefont {E.}~\bibnamefont {Bach}}, \bibinfo {author} {\bibfnamefont {M.}~\bibnamefont {Friesen}}, \bibinfo {author} {\bibfnamefont {R.}~\bibnamefont {Joynt}},\ and\ \bibinfo {author} {\bibfnamefont {S.~N.}\ \bibnamefont {Coppersmith}},\ }\href {https://doi.org/10.1103/PhysRevA.86.022334} {\bibfield  {journal} {\bibinfo  {journal} {Phys. Rev. A}\ }\textbf {\bibinfo {volume} {86}},\ \bibinfo {pages} {022334} (\bibinfo {year} {2012})}\BibitemShut {NoStop}%
\end{thebibliography}%

\end{document}